\RequirePackage{arydshln}
\documentclass[aps,twocolumn,nofootinbib,superscriptaddress,preprintnumbers,pra,10pt,floatfix]{revtex4-1}

\usepackage[utf8]{inputenc}

\usepackage{float}
\usepackage{amsmath,amssymb}
\usepackage{dsfont} 
\usepackage{hyperref}
\usepackage{graphicx}
\usepackage{enumitem}
\usepackage{mathtools}
\usepackage{bbold}
\usepackage{multirow}
\usepackage{ytableau}
\usepackage{youngtab}
\usepackage{braket}
\usepackage{soul}
\usepackage{cancel}
\usepackage{xcolor}
\usepackage[normalem]{ulem}
\usepackage{lipsum}

\usepackage{colortbl} 
\definecolor{Gray}{gray}{0.95}
\definecolor{RGray}{gray}{0.90}
\definecolor{CGray}{gray}{0.92}

\usepackage{arydshln}

\usepackage{tikz}
\usetikzlibrary{calc,tikzmark,fit,shapes.geometric,matrix,decorations.markings,arrows.meta,decorations.pathmorphing,patterns,positioning,snakes}

\tikzset
  {midarrow/.style={decoration={markings,mark=at position 0.5 with
     {\arrow[thin,xshift=2pt]{Triangle[length=4pt,#1]}}},postaction={decorate}}
  }

\tikzset{
proton/.style = {circle, draw=black, thin, fill=black!20!white, minimum size=#1,
              inner sep=0pt, outer sep=0pt},
proton/.default = 6pt 
}

\tikzset{
blob/.style = {circle, draw=black, thin, preaction={fill, black!20!white}, pattern=north east lines, minimum size=#1,
              inner sep=0pt, outer sep=0pt},
blob/.default = 6pt 
}

\tikzset{
wc/.style = {circle, fill, minimum size=#1,
              inner sep=0pt, outer sep=0pt},
wc/.default = 4pt 
}

\tikzset{vector/.style={decorate, decoration=snake}}

\makeatletter
\g@addto@macro\bfseries{\boldmath}
\makeatother

\makeatletter
\renewcommand\paragraph{\@startsection{paragraph}{4}{\z@}%
                                    {3.25ex \@plus1ex \@minus.2ex}%
                                    {-1em}%
                                    {\normalfont\normalsize\bfseries}}
\makeatother

\allowdisplaybreaks
\begin{document}


\title{Disentangling left and right-handed neutrino effects in $B\rightarrow K^{(*)}\nu\nu$}
 
\author{L.~P.~S.~Leal}
\email{luighi.leal@usp.br}
\affiliation{IJCLab, P\^ole Th\'eorie (Bat.~210), CNRS/IN2P3, 91405 Orsay, France}
\affiliation{Instituto de Física, Universidade de S\~ao Paulo, C.~P. 66.318, 05315-970 S\~ao Paulo, Brazil}
\author{S.~Rosauro-Alcaraz}
\email{salvador.rosauro@ijclab.in2p3.fr}
\affiliation{IJCLab, P\^ole Th\'eorie (Bat.~210), CNRS/IN2P3, 91405 Orsay, France}

\begin{abstract}
\vspace{5mm}
The first observation of $\mathcal{B}\left(B^+\rightarrow K^+\nu\nu\right)$ by the Belle II experiment lies almost $3\sigma$ away from the Standard Model expectation. In this letter we study this result in the SMEFT, extended by a light right-handed neutrino. We explore the correlations between the measured decay rate and other observables, such as $\mathcal{B}\left(B\rightarrow K^*\nu\nu\right)$ and $F_L\left(B\rightarrow K^*\nu\nu\right)$, showing that they could disentangle among scenarios involving left-handed neutrinos and those with the right-handed ones. Furthermore, we find that the high-$p_T$ tails of Drell-Yan processes studied at LHC provide important constraints that help us exclude some of the scenarios consistent with the Belle II result.
\vspace{3mm}
\end{abstract}

\maketitle

\allowdisplaybreaks

\section{Introduction}\label{sec:intro}
Flavor changing neutral current (FCNC) processes are one of the best arenas for observing indirect effects of new physics (NP), given that in the Standard Model (SM) they are only allowed at the loop level and further suppressed due to the unitarity of the CKM matrix. Thus, their measurement has the potential to reveal the presence of beyond the SM (BSM) states running in the loops. One such process is $b\rightarrow s ll$, where $l$ can either be a charged lepton or a neutrino. 

Over the past decade, the $B\rightarrow K^{(*)}\mu\mu$ decays were extensively studied in different experimental setups at LHC~\cite{LHCb:2014auh,LHCb:2015svh,LHCb:2016due,LHCb:2020lmf,CMS:2018qih}. The theoretical expectations for these observables are, however, hindered by large hadronic uncertainties due to the presence of long-distance effects involving $c\bar{c}$-resonances~\cite{Ciuchini:2015qxb}. This is not the case for the semileptonic decay into a pair of neutrinos, making it a much cleaner observable~\cite{Grossman:1995gt,Buchalla:1995vs,Buchalla:2000sk,Bartsch:2009qp,Buras:2014fpa} from the theoretical perspective.

Belle II recently reported on the evidence for $B^+\rightarrow K^++\mathrm{inv}$, with a branching fraction~\cite{Belle-II:2023esi}, 
\begin{equation}
    \mathcal{B}\left(B^+\rightarrow K^++\mathrm{inv}\right)^{\mathrm{exp}}=\left(2.3\pm 0.7\right)\times 10^{-5}\,.
\end{equation}
When compared with the SM prediction, for which recent results based on determinations of the relevant form factors on the lattice~\cite{Bailey:2015dka,Parrott:2022rgu} found~\cite{Becirevic:2023aov}
\begin{equation}
    \mathcal{B}\left(B^+\rightarrow K^+\nu\nu\right)^{\mathrm{SM}}=\left(4.44\pm 0.30\right)\times 10^{-6}\,,
\end{equation}
the experimental measurement is around $2.7\sigma$ above the SM result. The possibility of explaining this excess in terms of NP has been considered in several different scenarios~\cite{Athron:2023hmz,Bause:2023mfe,Allwicher:2023xba,He:2023bnk,Felkl:2023ayn,Altmannshofer:2023hkn,Dreiner:2023cms,Chen:2023wpb,Datta:2023iln,McKeen:2023uzo,Fridell:2023ssf,Ho:2024cwk,Chen:2024cll,Gabrielli:2024wys,Bolton:2024egx,He:2024iju,Chen:2024jlj,Hou:2024vyw}, ranging from its description within the SM effective field theory (EFT) assuming the NP particles to lie above the electroweak (EW) scale (see e.g. Refs.~\cite{Bause:2023mfe,Allwicher:2023xba}), to considerations involving the decay into some dark sector particle mimicking the neutrinos in the final state~\cite{Felkl:2023ayn,Altmannshofer:2023hkn,Hou:2024vyw,Bolton:2024egx}. 

Given the necessity of explaining the origin of neutrino masses, in this letter we will consider the possibility that one of the singlet neutrinos, necessary to generate neutrino masses in the Type-I seesaw and its variants~\cite{Minkowski:1977sc,Mohapatra:1979ia,Yanagida:1979as,Gell-Mann:1979vob,Branco:1988ex,Kersten:2007vk, Abada:2007ux,Mohapatra:1986aw,Mohapatra:1986bd,Akhmedov:1995ip,Malinsky:2005bi}, is light enough to play a role in the $b\rightarrow s\nu\nu$ transition, while any other NP effect is assumed to lie way above the EW scale.

We study the effect of all relevant dimension 6 four-fermion operators in the so-called $\nu$SMEFT~\cite{Liao:2016qyd}. Contributions from operators involving only the SM fields~\cite{Bause:2023mfe,Allwicher:2023xba} or those including the right-handed (RH) neutrino~\cite{Felkl:2023ayn} have been studied separately before. Here we will consider the possibility of having both types of operators, focusing on the synergies the study of different observables offers to tell apart between contributions with and without the RH neutrino. We find that, in some regions of the parameter space explaining the Belle II result, the measurement of the fraction of longitudinally polarized $K^*$ in the final state, $F_L$, allows to distinguish between RH neutrino contributions and those involving SM neutrinos. The importance of this observable in this regard has so far been overlooked in the context of this new measurement. Moreover, we consider the possibility of having a non-negligible mixing between the SM neutrinos and the heavy one. This in turn allows for the NP involving the RH neutrinos to indirectly affect the decay channels with light neutrinos, even when the decay to the heavy neutrino is kinematically forbidden. Finally, we profit from the interplay arising from the $SU(2)_L\times U(1)_Y$ symmetry to constrain some of the possible NP explanations of Belle II using processes involving charged leptons, with particular emphasis on the constraints that arise from the analysis of LHC data.

This paper is organised as follows: in Section~\ref{sec:nuSMEFT} we introduce the relevant effective operators that contribute to $b\rightarrow s\nu\nu$, and the connection with the low-energy EFT (LEFT) describing $B$-meson decays, including their relation to processes involving charged leptons in Sections~\ref{sec:LEFT} and~\ref{sec:Corr_chargedlept}. We discuss our results in Section~\ref{sec:Results}, and conclude in Section~\ref{sec:conclusions}.

\section{$\nu$SMEFT contributions to $b\rightarrow s\nu\nu$}\label{sec:nuSMEFT}
We extend the SM with $n_R$ RH neutrinos, $N_{sR}$ with $s=1,...,n_R$, at least one of which is light enough to participate in $B\rightarrow K^{(*)}\nu\nu$ decays, meaning $m_{n_4}\lesssim m_B-m_{K^{(*)}}$. The remaining sterile neutrinos, necessary to understand neutrino masses~\cite{Donini:2011jh}, as well as other BSM states, are considered to lie well above the EW scale. We can describe their effect on the $b\rightarrow s\nu\nu$ transition in terms of dimension $d=6$ operators as
\begin{equation}
    \mathcal{L}^{(6)}_{\nu\mathrm{SMEFT}}\supset \frac{1}{\Lambda^2}\sum_i \mathcal{C}_i\mathcal{O}_i\,,
\end{equation}
where $\Lambda$ is the new physics scale, and $\mathcal{C}_i$ is the Wilson coefficient (WC) for each $d=6$ operator, $\mathcal{O}_i$. There are only three four-femion operators involving only SM fields that can contribute to $B\rightarrow K^{(*)}\nu\nu$~\cite{Allwicher:2023xba} at tree level:
\begin{equation}
\begin{split}
    \left[\mathcal{O}^{(1)}_{lq}\right]_{\alpha\beta kl}&=\left(\bar{L}_\alpha \gamma^{\mu}L_\beta \right)\left(\bar{Q}_k \gamma_{\mu}Q_l\right)\,,\\
    \left[\mathcal{O}^{(3)}_{lq}\right]_{\alpha\beta kl}&=\left(\bar{L}_\alpha \gamma^{\mu}\tau^IL_\beta \right)\left(\bar{Q}_k\tau^I\gamma_{\mu}Q_l\right)\,,\\
    \left[\mathcal{O}_{ld}\right]_{\alpha\beta kl}&=\left(\bar{L}_\alpha \gamma_{\mu}L_\beta \right)\left(\bar{d}_{kR}\gamma^{\mu}d_{lR}\right)\,,
\end{split}
\label{eq:Op_SMEFT}
\end{equation}
where $L$ and $Q$ are the lepton and quark $SU(2)_L$ doublets, respectively, and $d_R$ are the down-type quark weak singlets. The Pauli matrices are denoted by $\tau^I$, with $I=1,\,2,\,3$. In the following, we will work in the flavor basis defined by the diagonal down-type quark Yukawa matrix, with the CKM matrix element in the upper component of $Q_i=\begin{pmatrix}\left(V^{\dagger}u\right)_i & d_i\end{pmatrix}^T$. Moreover, we will fix the quark flavor indexes to $k\,(l)=2,3$, relevant to $b\rightarrow s\nu\nu$, while $\alpha\,(\beta)\in\lbrace e,\,\mu,\,\tau\rbrace$. 

When considering the RH neutrino field, $N_R$, additional $d=6$ operators can contribute to $b\rightarrow s\nu\nu$~\cite{Liao:2016qyd,Felkl:2021uxi,Felkl:2023ayn}.~\footnote{Note that, in the context of the SMEFT, scalar and tensor contributions can arise from $d=7$ lepton number violating operators~\cite{Marzocca:2024hua} or $d=8$ lepton number conserving ones~\cite{Alonso:2014csa}.} Focusing on the tree-level contributions to this process, the relevant operators are:
\begin{equation}
\begin{split}
	\left[\mathcal{O}_{Nq}\right]_{kl}&=\left(\bar{N}_R\gamma^{\mu}N_R\right)\left(\bar{Q}_k \gamma_{\mu}Q_l\right)\,,\\
	\left[\mathcal{O}_{Nd}\right]_{kl}&=\left(\bar{N}_R\gamma^{\mu}N_R\right)\left(\bar{d}_{kR}\gamma^{\mu}d_{lR}\right)\,,\\
	\left[\mathcal{O}^S_{lNqd}\right]_{\alpha kl}&=\left(\bar{L}_\alpha  N_R\right)\epsilon \left(\bar{Q}_kd_{lR}\right)\,,\\
	\left[\mathcal{O}^T_{lNqd}\right]_{\alpha kl}&=\left(\bar{L}_\alpha  \sigma_{\mu\nu}N_R\right)\epsilon \left(\bar{Q}_k\sigma^{\mu\nu}d_{lR}\right)\,,
\end{split}
\label{eq:Op_nuSMEFT}
\end{equation}
where $\epsilon_{12}=1$. Compared to the SMEFT case, one can now have scalar and tensor operators contributing to $B\rightarrow K^{(*)}\nu\nu$. In the following, in order to ease the notation, we will not explicitly show the quark family indexes for the vector operators. On the other hand, for the scalar and tensor operators we keep them. This is because, depending on the generation index that we assign to the $SU(2)_L$ quark doublet, the $b\rightarrow s\nu\nu$ process will be correlated to different observables. 

Moreover, the $SU(2)_L\times U(1)_Y$ symmetry will allow to relate processes involving neutrinos with those involving charged leptons in the final state~\cite{Bause:2020auq,Bause:2021cna}. This was studied in the context of SMEFT in Refs.~\cite{Bause:2023mfe,Allwicher:2023xba}. We will also consider the effect of the $\nu$SMEFT operators from Eq.~(\ref{eq:Op_nuSMEFT}) onto $b\rightarrow c\ell \nu$, whose impact on $B\rightarrow D^{(*)}\ell \nu$ was discussed in Ref.~\cite{Felkl:2023ayn}, but neglecting the neutrino mixing effects. We will also study how the LHC data can help us further constrain couplings to the  four-fermion operators listed above, with particular emphasis on $\mathcal{O}^{S(T)}_{lNqd}$. 

In the following, we will study the impact that the WCs, computed at high-energies, have on low-energy observables. We will need to take into account running effects for the scalar and tensor operators, which do not mix in QCD. To account for this running and be able to consistently compare bounds on the WCs arising at low-energies with those related to searches at LHC, we will consider QCD running at 3-loop level~\cite{Gonzalez-Alonso:2017iyc}. Running effects from the TeV to the $m_b$ scale at 3-loop order can differ from those at 1-loop level by up to $10\%$. Whenever we quote limits or allowed regions for the WCs, we systematically compute them at the energy scale $\mu=1$~TeV.

\section{Low-energy implications for $b\rightarrow s\nu\nu$}\label{sec:LEFT}
Neutrino masses are generated after spontaneous symmetry breaking. On general grounds, the neutrino flavor eigenstates, $\nu_{\alpha L}$ and $N_R$, are an admixture of the mass eigenstates, $n_i$, which in the following we consider to be Majorana states satisfying $n_i^c=n_i$. These two bases are related through a unitary matrix, $\mathcal{U}$, such as
\begin{equation}
\begin{split}
	\nu_{\alpha L}&=\sum_{i=1}^4 \mathcal{U}_{\alpha i}P_Ln_i\,,\quad \mathrm{for}\,\alpha=e,\,\mu,\,\tau\,,\\
	N_R&=\sum_{i=1}^4 \mathcal{U}_{s i}^* P_R n_i\,,
\end{split}
\label{eq:nu_mixing}
\end{equation}
where the sum over mass eigenstates only runs through the states which are relevant to $b\rightarrow s\nu\nu$, and $P_{L(R)}=(1\mp \gamma_5)/2$.~\footnote{We refer the reader to Appendix~\ref{app:Neutrino_masses} for more details.} Even though the current experimental limits constrain the active-heavy mixing to be $|\mathcal{U}_{\alpha 4}|\ll 1$, these bounds vary depending both on the mass scale of the heavy neutrino, $m_{n_4}$, and on the flavor of the active neutrino with which it primarily mixes. We will consider the effect of non-zero mixing with the $\tau$ neutrino~\cite{Fernandez-Martinez:2023phj}, while neglecting mixing with the electron or muon neutrinos, $|\mathcal{U}_{e(\mu)4}|\simeq 0$. 

When considering the LEFT with massive neutrinos, it proves advantageous to describe $b\rightarrow s\nu\nu$ in terms of operators with well-defined parity properties. Similarly to the $b\rightarrow s\ell^+_\alpha \ell^-_{\beta}$ case~\cite{Altmannshofer:2008dz,Altmannshofer:2009ma,Buras:2014fpa,Gratrex:2015hna,Becirevic:2016zri}, the low-energy lagrangian is given by
\begin{equation}
	\mathcal{L}^{b\rightarrow s\nu\nu}_{\mathrm{LEFT}}\supset \frac{4G_F}{\sqrt{2}}\lambda_t \frac{\alpha_{\mathrm{em}}}{8\pi}\sum_{a=V,A,S,P,T}\left[C_a\mathcal{O}_a+C^{\prime}_a\mathcal{O}^{\prime}_a\right]+h.c.\,,
\label{eq:Lag_LEFT}
\end{equation}
where $G_F$ is the Fermi constant, $\alpha_{\mathrm{em}}$ the fine structure constant and $\lambda_t\equiv V_{ts}^*V_{tb}$ is the combination of CKM matrix elements relevant to the $b\rightarrow s$ transition. The extra factor of 2, with respect to the usual normalization, arises from the Majorana nature of neutrinos, and allows one to directly use the results from Ref.~\cite{Gratrex:2015hna}, as discussed in Ref.~\cite{Felkl:2021uxi}. The operators in this basis are
\begin{equation}
\begin{split}
	\left[\mathcal{O}_{V(A)}\right]_{ij} &= \left(\bar{s}_L\gamma_{\mu}b_L\right) \left(\bar{n}_i\gamma^{\mu}(\gamma_5)n_j\right)\,,\\
	\left[\mathcal{O}_{S(P)}\right]_{ij} &= \left(\bar{s}_L b_R\right) \left(\bar{n}_i(\gamma_5)n_j\right)\,,\\
	\left[\mathcal{O}_T\right]_{ij} & = \left(\bar{s}_L\sigma_{\mu\nu}b_R\right)\left(\bar{n}_i\sigma^{\mu\nu}n_j\right)\,,
\end{split}
\label{eq:Op_LEFT}
\end{equation}
while primed operators, $\mathcal{O}^{\prime}_a$, are obtained by consistently exchanging $q_L\leftrightarrow q_R$ in Eq.~(\ref{eq:Op_LEFT}), with $q\in\lbrace s,\,b\rbrace $.~\footnote{In Eq.~(\ref{eq:Lag_LEFT}) the sum over neutrino states, $i,j$, is not shown explicitely.} We note here that the WCs  $C^{(\prime)}_a$ have well-defined symmetry properties under the exchange of neutrino indexes, assuming they are Majorana particles. It can be shown that the WCs related to (pseudo-) scalar and axial-vector operators are symmetric in the neutrino mass-eigenstate indexes, $i, j$. The vector and tensor WCs, instead, are antisymmetric with respect to the exchange $i\leftrightarrow j$. This in turn implies that vector and tensor operators will be neutrino flavor violating by construction, meaning that the decay $B\rightarrow K^{(*)}\nu_i \nu_j$ will always have $i\neq j$. 

Taking into account the neutrino mixing matrix, $\mathcal{U}$, we find the following relation between the WCs in the LEFT and those from the $\nu$SMEFT operators listed in Eq.~(\ref{eq:Op_nuSMEFT}):
\begin{equation}
\begin{split}
	\left[C_{S(P)}\right]_{ij}&=\mathcal{G}\sum_{\alpha} \left[\mathcal{C}^S_{lNqd}\right]_{\alpha 23}\left(\mathcal{U}_{\alpha i}^*\mathcal{U}_{s j}^*+\mathcal{U}_{\alpha j}^*\mathcal{U}_{s i}^*\right)\,,\\
	\left[C^{\prime}_{S(P)}\right]_{ij}&=\pm\mathcal{G}\sum_{\alpha} \left[\mathcal{C}^S_{lNqd}\right]^*_{\alpha 32}\left(\mathcal{U}_{\alpha i}\mathcal{U}_{s j}+\mathcal{U}_{\alpha j}\mathcal{U}_{s i}\right)\,,
\end{split}
\label{eq:Matching_LEFT_nuSMEFT_Scalar}
\end{equation}
where the prefactor is defined as $\mathcal{G}\equiv \pi v^2/(\Lambda^2\lambda_t\alpha_{\mathrm{em}})$ with $v\simeq246.22$~GeV, and the sum runs over $\alpha\in \lbrace e,\,\mu,\,\tau\rbrace$. For the vector and axial WCs we get
\begin{equation}
\begin{split}
	\left[C_V\right]_{ij}&=-2i \mathcal{G}\left[\mathcal{C}_{Nq}\right]\sum_{\alpha}\mathrm{Im}\left[\mathcal{U}_{\alpha i}^*\mathcal{U}_{\alpha j}\right]\,,\\
	\left[C^{\prime}_V\right]_{ij}&=-2i \mathcal{G}\left[\mathcal{C}_{Nd}\right]\sum_{\alpha}\mathrm{Im}\left[\mathcal{U}_{\alpha i}^*\mathcal{U}_{\alpha j}\right]\,,\\
	\left[C_A\right]_{ij}&=2\mathcal{G}\left[\mathcal{C}_{Nq}\right]\left(\delta_{ij}-\sum_{\alpha}\mathrm{Re}\left[\mathcal{U}_{\alpha i}^*\mathcal{U}_{\alpha j}\right]\right)\,,\\
 \left[C^{\prime}_A\right]_{ij}&=2\mathcal{G}\left[\mathcal{C}_{Nd}\right]\left(\delta_{ij}-\sum_{\alpha}\mathrm{Re}\left[\mathcal{U}_{\alpha i}^*\mathcal{U}_{\alpha j}\right]\right)\,.
\end{split}
\label{eq:Matching_LEFT_nuSMEFT_Vector}
\end{equation}
Finally, for the tensor WCs we find
\begin{equation}
\begin{split}
    \left[C_T\right]_{ij}&=2\mathcal{G}\sum_{\alpha}\left[\mathcal{C}^T_{lNqd}\right]_{\alpha 23}\left(\mathcal{U}_{\alpha i}^*\mathcal{U}_{s j}^*-\mathcal{U}_{\alpha j}^*\mathcal{U}_{s i}^*\right)\,,\\
	\left[C^{\prime}_T\right]_{ij}&=-2\mathcal{G}\sum_{\alpha}\left[\mathcal{C}^T_{lNqd}\right]^*_{\alpha 32}\left(\mathcal{U}_{\alpha i}\mathcal{U}_{s j}-\mathcal{U}_{\alpha j}\mathcal{U}_{s i}\right)\,.
\end{split}
\label{eq:Matching_LEFT_nuSMEFT_Tensor}
\end{equation}
Clearly, in the presence of a single RH neutrino, $C_V^{(\prime)}$ in Eq.~(\ref{eq:Matching_LEFT_nuSMEFT_Vector}) can contribute to $b\rightarrow s\nu\nu$ only if there are sources of CP violation in the neutrino sector. Note also that when relating the WCs in the LEFT to those in the $\nu$SMEFT one finds $C_P=C_S$ and $C^{\prime}_P=-C_S^{\prime}$. 

Operators involving only SM fields (see Eq.~(\ref{eq:Op_SMEFT})) can also contribute to $C_V^{(\prime)}$ and $C_A^{(\prime)}$. These contributions are given by
\begin{equation}
\begin{split}
	\left[C_{V(A)}\right]_{ij}&=\pm\mathcal{G}\sum_{\alpha,\beta} \left[\mathcal{C}_{lq}^{(1)}+\mathcal{C}_{lq}^{(3)}\right]_{\alpha\beta}\left(\mathcal{U}_{\alpha i}^*\mathcal{U}_{\beta j}\mp\mathcal{U}_{\alpha j}^*\mathcal{U}_{\beta i}\right)\,,\\
	\left[C^{\prime}_{V(A)}\right]_{ij}&=\pm\mathcal{G}\sum_{\alpha,\beta} \left[\mathcal{C}_{ld}\right]_{\alpha\beta}\left(\mathcal{U}_{\alpha i}^*\mathcal{U}_{\beta j}\mp\mathcal{U}_{\alpha j}^*\mathcal{U}_{\beta i}\right)\,,
\end{split}
\label{eq:Matching_LEFT_SMEFT}
\end{equation}
where $\alpha,\beta\in \lbrace e,\,\mu,\,\tau\rbrace$. In the following we will study the effect these operators have on $B\rightarrow K^{(*)}\nu\nu$ for different mass ranges of the sterile state. Note that, even if the heavy neutrino is not kinematically accesible in the $B$ decay, its mixing with the SM neutrinos could nonetheless produce an indirect effect on these observables. 

The branching fraction for $B\rightarrow K^{(*)}\nu\nu$ in the presence of NP can be written as
\begin{equation}
    \mathcal{B}\left(B\rightarrow K^{(*)}\nu\nu\right)=\mathcal{B}\left(B\rightarrow K^{(*)}\nu\nu\right)^{\mathrm{SM}}\left(1+\delta\mathcal{B}_{K^{(*)}}\right),
\end{equation}
where all the effects of NP are encoded in $\delta \mathcal{B}_{K^{(*)}}$.~\footnote{Notice that the WCs entering $\delta \mathcal{B}_{K^{(*)}}$ are evaluated at the scale of the process, $\mu=m_b$.} Whenever $m_{n_4}\ll m_B-m_K$ so that the sterile mass is negligible, the deviations from the SM prediction can be easily written as
\begin{equation}
	\begin{split}
    		\delta\mathcal{B}_{K}=&-\frac{\sum_{i,j}^3\mathrm{Re}\left[C_L^{\mathrm{SM}}\delta_{ij}\left(C_A+C_A^{\prime}\right)_{ij}\right]}{3|C_L^{\mathrm{SM}}|^2}\\
		+&\frac{1}{12|C_L^{\mathrm{SM}}|^2}\sum_{i,j}^4\bigg\lbrace |C_V+C_V^{\prime}|_{ij}^2+|C_A+C_A^{\prime}|_{ij}^2\\
		+&\frac{3}{2}\eta_K^S\left(|C_S+C_S^{\prime}|_{ij}^2+|C_P+C_P^{\prime}|_{ij}^2\right)\\
		+&4\eta_K^T\left(|C_T|^2_{ij}+|C_T^{\prime}|_{ij}^2\right)\bigg\rbrace\,,
	\end{split}
	\label{eq:deltaBK}
\end{equation}
where $C_L^{\mathrm{SM}}=-6.32(7)$ is the SM contribution to $b\rightarrow s\nu\nu$.~\footnote{We refer the reader to Appendix~\ref{sec:diff_dec_rates} for the general expressions. Note that the SM contribution to $b\rightarrow s\nu\nu$ is flavor diagonal and universal.} In the case of $B\rightarrow K^*\nu\nu$, neglecting once again the sterile neutrino mass, we find
\begin{equation}
    \begin{split}
        \delta\mathcal{B}_{K^*}=&-\frac{\sum_{i,j}^3 \mathrm{Re}\left[C_L^{\mathrm{SM}}\delta_{ij}(C_A+C_A^{\prime})_{ij}\right]}{3|C_L^{\mathrm{SM}}|^2}\\
        +&\eta^V_{K^*}\frac{\sum_{i,j}^3 \mathrm{Re}\left[C_L^{\mathrm{SM}}\delta_{ij}(C_A^{\prime})_{ij}\right]}{6|C_L^{\mathrm{SM}}|^2}\\
        +&\frac{1}{12|C_L^{\mathrm{SM}}|^2}\sum_{i,j}^4\bigg\lbrace \frac{\eta^V_{K^*}}{4}\left(|C_V-C_V^{\prime}|_{ij}^2+|C_A-C_A^{\prime}|_{ij}^2\right)\\
        +&\left(1-\frac{\eta^V_{K^*}}{4}\right)\left(|C_V+C_V^{\prime}|_{ij}^2+|C_A+C_A^{\prime}|_{ij}^2\right)\\
        +&\frac{4}{3}\eta^S_{K^*}\left(|C_S-C_S^{\prime}|_{ij}^2+|C_P-C_P^{\prime}|_{ij}^2\right)\\
        +&4\eta^T_{K^*}\left(|C_T|^2_{ij}+|C_T^{\prime}|_{ij}^2\right)\bigg\rbrace\,.
    \end{split}
    \label{eq:deltaBKst}
\end{equation}
Note that in Eqs.~(\ref{eq:deltaBK}-\ref{eq:deltaBKst}), the SM contribution only interferes with that of the axial-vector operators, $\mathcal{O}_A^{(\prime)}$, involving light neutrinos, $n_i$ with $i=1,2,3$.~\footnote{When including the mixing between the active and the heavy neutrinos, there is a small non-diagonal contribution proportional to $\mathcal{U}_{s i}^*\mathcal{U}_{s j}$. This would allow for the interference between SM and the $\nu$SMEFT axial-vector operators involving an $N_R$, but it is nonetheless mixing suppressed and negligible in comparison with the other NP effects arising directly from the $d=6$ operators.} The prefactors $\eta_{K^{(*)}}^a$ depend solely on the hadronic form factors, and we find the following numerical values:
\begin{equation}
	\begin{split}
       &\eta_K^S=0.940(16)\,,\,  \eta_K^T=0.270(11)\,, \\
        	&\eta_{K^*}^S=0.34(2)\,,\,   \eta_{K^*}^V=3.33(5)\,,\,  \eta_{K^*}^T=1.58(16)\,. 
\end{split}
\label{eq:values_etas} 
\end{equation}
Explicit expressions in terms of the hadronic form factors are provided in Appendix~\ref{sec:Form_factors}. 

If the Belle II result is to be described only by SMEFT operators, a lower bound on $\mathcal{B}\left(B\rightarrow K^*\nu\nu\right)$ can be deduced from $\mathcal{B}\left(B\rightarrow K\nu\nu\right)$. This is given by~\cite{Bause:2023mfe,Allwicher:2023xba}
\begin{equation}
    \frac{\mathcal{B}\left(B\rightarrow K^*\nu\nu\right)}{\mathcal{B}\left(B\rightarrow K\nu\nu\right)}\geq \left(1-\frac{\eta_{K^*}^V}{4}\right)\frac{\mathcal{B}\left(B\rightarrow K^*\nu\nu\right)^{\mathrm{SM}}}{\mathcal{B}\left(B\rightarrow K\nu\nu\right)^{\mathrm{SM}}}.
    \label{eq:bound_EFT}
\end{equation}
We find that this inequality is still satisfied in the $\nu$SMEFT for the (axial-) vector and tensor operators, whenever the heavy neutrino mass can be neglected, $m_{n_4}\ll m_B-m_{K}$. 

This, however, is not the case for the (pseudo-) scalar operators, or when $m_{n_4}$ is large. As can be understood from Eq.~(\ref{eq:deltaBKst}), one could consider the possibility that $C_S=C_S^{\prime}$ and $C_P=C_P^{\prime}$ in the absence of other NP contributions.~\footnote{This situation is, however, not possible with the effective operators we have considered in Eq.~\ref{eq:Op_nuSMEFT}, but could arise in more general scenarios.} In this limit one would have $\delta\mathcal{B}_{K^*}=0$ while having $\delta\mathcal{B}_K\neq 0$. Under such circumstances, the inequality in Eq.~(\ref{eq:bound_EFT}) would be violated provided that 
\begin{equation}
    \mathcal{B}\left(B\rightarrow K\nu\nu\right)>\frac{4\mathcal{B}\left(B\rightarrow K\nu\nu\right)^{\mathrm{SM}}}{4-\eta_{K^*}^V}\,,
\end{equation}
for which we find numerically $\mathcal{B}\left(B\rightarrow K\nu\nu\right)\gtrsim 3.1\times 10^{-5}$. Moreover, when $m_{n_4}$ cannot be neglected, Eq.~(\ref{eq:bound_EFT}) does not apply. Indeed, it is always possible to reduce the available phase space for the $B\rightarrow K^*\nu\nu$ decay into the heavy neutrino, recovering the SM result, while simultaneously having a non-negligible contribution to $B\rightarrow K\nu\nu$ from NP.~\footnote{This is only strictly true in the absence of neutrino mixing.}

NP described by the lagrangian in Eq.~(\ref{eq:Lag_LEFT}) can be relevant to $B_s\rightarrow \nu\nu$, for which an upper bound $\mathcal{B}\left(B_s\rightarrow \nu\nu\right)<5.9\times 10^{-4}$ at 90\% C. L. was obtained using LEP data~\cite{Alonso-Alvarez:2023mgc}. We can write its branching fraction as~\cite{Becirevic:2016zri}:
\begin{equation}
	\begin{split}
		\mathcal{B}\big(B_s&\rightarrow \nu\nu\big)=\frac{\tau_{B_s}}{128\pi^3}\frac{G_F^2|\lambda_t|^2\alpha_{\mathrm{em}}^2}{m_{B_s}^3}f_{B_s}^2\sum_{i,j=1}^4\lambda_{B_sn_in_j}^{1/2}\\
		\bigg\lbrace &\left[m_{B_s}^2-(m_i+m_j)^2\right]\bigg|(C_V-C_V^{\prime})_{ij}(m_i-m_j)\\
		+&(C_S-C_S^{\prime})_{ij}\frac{m_{B_s}^2}{m_b+m_s}\bigg|^2+\left[m_{B_s}^2-(m_i-m_j)^2\right]\\
		\bigg|&(C_A-C_A^{\prime})_{ij}(m_i+m_j)+(C_P-C_P^{\prime})_{ij}\frac{m_{B_s}^2}{m_b+m_s}\bigg|^2\bigg\rbrace,
	\end{split}
	\label{eq:BR_BstoInv}
\end{equation}
where $\lambda_{abc}=\left[m_a^2-(m_b-m_c)^2\right]\left[m_a^2-(m_b+m_c)^2\right]$. From Eq.~(\ref{eq:BR_BstoInv}) it is clear that the contribution from the (axial-) vector operators will be suppressed by the tiny neutrino masses, except for the heavy neutrino. On the contrary, the (pseudo-) scalar operator, which is not chirality suppressed, can considerably enhance $\mathcal{B}\left(B_s\rightarrow \nu\nu\right)$. This channel then proves a complementary probe to constrain the NP explanations of the Belle II result.

Finally, by interpreting the size of the WC in terms of a NP coupling constant, $g_{\mathrm{NP}}$, and the NP scale, $\Lambda$, related to a mediator mass, i.e. $\mathcal{C}/\Lambda^2\simeq g_{\mathrm{NP}}^2/\Lambda^2$, we can estimate the maximal cutoff where the NP needs to lie in order to accommodate the Belle II result while having perturbative couplings $g_{\mathrm{NP}}\lesssim \sqrt{4\pi}$~\cite{Farina:2016rws}. We find that for operators involving the SM neutrinos, $\Lambda \lesssim 24$~TeV, while in the case of scalar operators we find a somewhat larger scale, $\Lambda \lesssim 32$~TeV.

\section{Correlation with processes involving charged leptons}\label{sec:Corr_chargedlept}
When considering NP described by the $\nu$SMEFT, correlations between processes involving neutrinos and those with charged leptons arise thanks to the $SU(2)_L$ gauge symmetry. All the operators from Eq.~(\ref{eq:Op_SMEFT}) contribute to $b\rightarrow s\ell_{\alpha} ^+\ell_{\beta}^-$, which implies that the most plausible explanation for the Belle II result within the SMEFT is that NP couples to the third generation leptons~\cite{Allwicher:2023xba}, unless there is some ad-hoc cancellation between various contributions. 
Moreover, the operators from Eqs.~(\ref{eq:Op_SMEFT}-\ref{eq:Op_nuSMEFT}) relate the $b\rightarrow s\nu\nu$ transition measured by Belle II to the semileptonic $b\rightarrow c\ell_{\alpha}\nu$ process. At low-energies, assuming NP described by the same $\nu$SMEFT operators that enter $b\rightarrow s\nu\nu$, we can write
\begin{equation}
	\begin{split}
		\mathcal{L}^{b\rightarrow c\ell \nu}_{\mathrm{LEFT}}\supset \sqrt{2}G_FV_{cb}\sum_{a=V,A,S,P,T}\left[C_a \mathcal{O}_a+C_a^{\prime}\mathcal{O}_a^{\prime}\right]+h.c.,
	\end{split}
	\label{eq:Lag_LEFT_btoc}
\end{equation}
where the only operators arising from $\nu$SMEFT are
\begin{equation}
	\begin{split}
	\left[\mathcal{O}_{V(A)}\right]_{\alpha i}&=\left(\bar{c}_L\gamma_{\mu}b_L\right)\left(\bar{\ell}_{\alpha}\gamma^{\mu}(\gamma_5)n_i\right),\\
	\left[\mathcal{O}_{S(P)}\right]_{\alpha i}&=\left(\bar{c}_Lb_R\right)\left(\bar{\ell}_{\alpha} (\gamma_5)n_i\right),\\
	\left[\mathcal{O}_T\right]_{\alpha i}&=\left(\bar{c}_L\sigma_{\mu\nu}b_R\right)\left(\bar{\ell}_{\alpha}\sigma^{\mu\nu}n_i\right).
	\end{split}
	\label{eq:Op_b_to_c}
\end{equation}
Although the chirality flipped operators with a RH charm quark, $\mathcal{O}_a^{\prime}$, can in general be generated within the ($\nu$)SMEFT, they are not related to the relevant operators interesting for $b\rightarrow s\nu\nu$, and thus we do not consider them in the following. The relation between WCs at low energies and the $\nu$SMEFT ones is given by
\begin{equation}
	\begin{split}
		\left[C_{V(A)}^{b\rightarrow c\ell\nu}\right]_{\alpha i}=&C_{V(A)}^{\mathrm{SM}}\mathcal{U}_{\alpha i}\pm\frac{v^2}{\Lambda^2}\frac{V_{cs}}{V_{cb}}\sum_{\beta}\left[\mathcal{C}_{lq}^{(3)}\right]_{\alpha \beta}\mathcal{U}_{\beta i},\\
		\left[C_{S(P)}^{b\rightarrow c\ell\nu}\right]_{\alpha i}=&- \frac{v^2}{2\Lambda^2}\frac{V_{cs}}{V_{cb}}\left[\mathcal{C}_{lNqd}^S\right]_{\alpha 23}\mathcal{U}^*_{si},\\
		\left[C_T^{b\rightarrow c\ell\nu}\right]_{\alpha i}=&- \frac{v^2}{\Lambda^2}\frac{V_{cs}}{V_{cb}}\left[\mathcal{C}_{lNqd}^T\right]_{\alpha 23}\mathcal{U}^*_{si},
	\end{split}
	\label{eq:WC_b_to_c}
\end{equation}
where $C_{V(A)}^{\mathrm{SM}}=\pm 1$ corresponds to the SM.~\footnote{Note that operators like $\left[\mathcal{O}_{lNqd}^{S(T)}\right]_{\alpha 32}$ would contribute to semileptonic transitions with top quarks.} In order to better understand the interplay between the WCs relevant to $b\rightarrow s\nu\nu$ and those entering $b\rightarrow c\ell_{\alpha}\nu$, it is useful to write the latter in terms of those from Eqs.~(\ref{eq:Matching_LEFT_nuSMEFT_Scalar}-\ref{eq:Matching_LEFT_SMEFT}).~\footnote{We assume here that a single operator is responsible simultaneously for the NP effects in both processes.} We find:
\begin{equation}
	\begin{split}
        \left[C_{V(A)}^{b\rightarrow c\ell\nu}\right]_{\alpha i}=&C_{V(A)}^{\mathrm{SM}}\mathcal{U}_{\alpha i}\mp \frac{\alpha_{\mathrm{em}}\lambda_t}{2\pi}\frac{V_{cs}}{V_{cb}}\\
        &\sum_{j}^4\mathcal{U}_{\alpha j}\left(\left[C_V\right]_{ij}+\left[C_A\right]_{ij}\right)\,,\\
		\left[C_{S(P)}^{b\rightarrow c\ell\nu}\right]_{\alpha i}=&-\frac{V_{cs}}{V_{cb}}\frac{\alpha_{\mathrm{em}}\lambda_t}{2\pi}\sum_{j}^4\left[C_{S}\right]_{ij}\mathcal{U}_{\alpha j}\,,\\
        \left[C_{T}^{b\rightarrow c\ell\nu}\right]_{\alpha i}=&\frac{V_{cs}}{V_{cb}}\frac{\alpha_{\mathrm{em}}\lambda_t}{\pi}\sum_{j}^4\left[C_{T}\right]_{ij}\mathcal{U}_{\alpha j}\,,
	\end{split}
 \label{eq:rel_btoc_btos}
\end{equation}
It is then clear from Eq.~(\ref{eq:rel_btoc_btos}) that, even if we have a non-negligible NP effect in $b\rightarrow s\nu\nu$ from, say, $C_{S(P)}\neq 0$, its corresponding effect in $b\rightarrow c\ell \nu$ will be suppressed by $\alpha_{\mathrm{em}}\lambda_t/2\pi \sim 5\times 10^{-5}$, so that only large WCs can have an impact in $R_{D^{(*)}}$. This is also related to the fact that $b\rightarrow c\ell\nu$ arises at tree-level in the SM, in contrast to $b\rightarrow s\nu\nu$ which is not only loop but also CKM-suppressed.

It is also useful to account for the constraints on such $d=6$ operators from collider searches. In particular, we can use the high-$p_T$ tails of Drell-Yan processes involving a charged lepton and missing energy in the final state, $pp\rightarrow \ell_{\alpha}\nu$, in order to constrain the size of the scalar and tensor WCs~\cite{Farina:2016rws,Allwicher:2022gkm,Allwicher:2022mcg}. As we will see, these searches probe a good part of the parameter space explaining $\mathcal{B}\left(B\rightarrow K\nu\nu\right)^{\mathrm{exp}}$, mostly for large heavy neutrino masses.

\section{Results}\label{sec:Results}
\subsection{$b\rightarrow s\nu\nu$}
In the following, we discuss the impact of the $d=6$ NP operators relevant to $b\rightarrow s\nu\nu$ for different assumptions on the heavy neutrino mass and on its mixing with the SM neutrinos. In all instances in which lepton doublets are involved, we introduce the NP only in the $\tau$-sector. This is motivated for operators involving only SM fields by the strong constraints coming from $\mathcal{B}\left(B_s\rightarrow \mu\mu\right)$ and $R_{K^{(*)}}\equiv \mathcal{B}\left(B\rightarrow K^{(*)} \mu\mu\right)/\mathcal{B}\left(B\rightarrow K^{(*)} ee\right)$, which rule out the possibility to accommodate $\mathcal{B}\left(B\rightarrow K\nu\nu\right)^{\mathrm{exp}}$ with couplings to light charged leptons~\cite{Bause:2023mfe,Allwicher:2023xba}.
\begin{figure*}
\centering
    \includegraphics[width=0.45\textwidth]{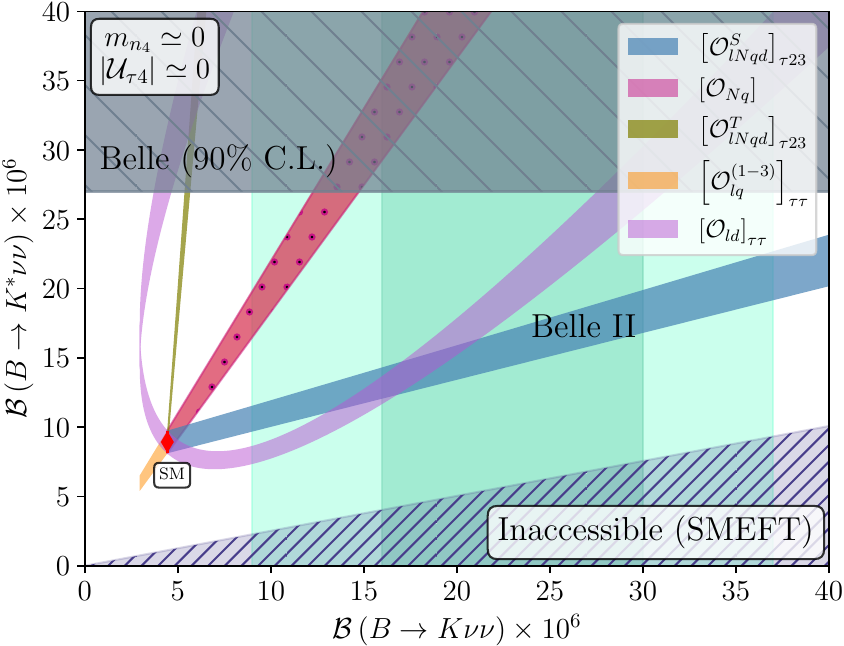}
    \includegraphics[width=0.45\textwidth]{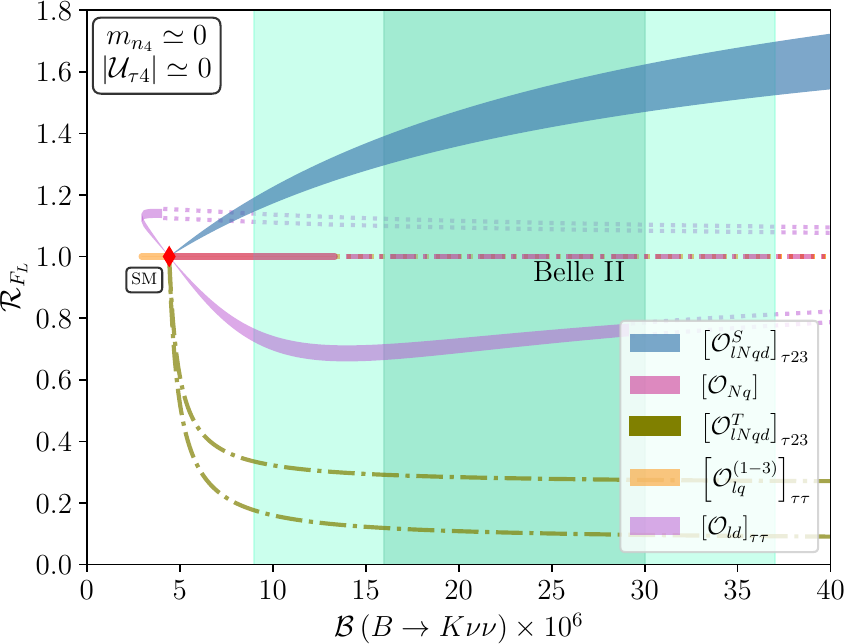}
    \caption{\small \sl Correlation between $\mathcal{B}\left(B\rightarrow K\nu\nu\right)$ and $\mathcal{B}\left(B\rightarrow K^*\nu\nu\right)$ (left panel) when including the effect of one single NP operator at a time and $m_{n_4}\ll m_B-m_K$. The red point corresponds to the SM result and the (light) green regions to the Belle II result at $1(2)\sigma$. The hatched areas on the left panel correspond to the Belle upper bound, $\mathcal{B}\left(B\rightarrow K^*\nu\nu\right)<2.5\times 10^{-5}$, in gray, and the the region described by Eq.~(\ref{eq:bound_EFT}) (blue area). In the right panel we show the correlation with $\mathcal{R}_{F_L}=F_L/F_L^{\mathrm{SM}}$. We show, as solid bands, the regions which satisfy the Belle upper bound on $\mathcal{B}\left(B\rightarrow K^*\nu\nu\right)$, while as (dash-) dotted lines those excluded.}
    \label{fig:BR_BtoK_vs_BR_BtoKst}
\end{figure*}
\subsubsection{No neutrino mixing}
\label{sec:NoNuMix}
Let us first consider the effect of NP in the absence of mixing, $\mathcal{U}_{\tau 4}=0$. In this case there is no interference between the SM and the effective operators involving a RH neutrino, cf. Eq.~(\ref{eq:Op_nuSMEFT}). Therefore, the branching fraction for $B\rightarrow K^{(*)}\nu\nu$ in the presence of this NP can only be larger than the SM expectation. Furthermore, the effect of the operator $\mathcal{O}_{Nd}$ is degenerate with $\mathcal{O}_{Nq}$. Likewise, $\left[\mathcal{O}^{S(T)}_{lNqd}\right]_{\tau 32}$  has the same impact as $\left[\mathcal{O}^{S(T)}_{lNqd}\right]_{\tau 23}$.
Contrary, operators with only LH neutrinos, $\mathcal{O}_{lq}^{(1-3)}$ and $\mathcal{O}_{ld}$, do interfere with the SM. Nonetheless, given that we are introducing NP only in the $\tau$-sector, the destructive interference between SM and NP effects cannot be exact and thus it is not possible to arbitrarily suppress $\mathcal{B}\left(B\rightarrow K^{(*)}\nu\nu\right)$.

In the following we study these contributions for two representative mass ranges for the heavy neutrino. 
\begin{itemize}
    \item $m_{n_4}\ll m_B-m_K$
    
When we neglect the heavy neutrino mass, $\mathcal{O}^{(1-3)}_{lq}$, shown in orange in Fig.~\ref{fig:BR_BtoK_vs_BR_BtoKst}, has the same effect as $\mathcal{O}_{Nq}$ (shown in hatched pink) on the region $\mathcal{B}\left(B\rightarrow K^{(*)}\nu\nu\right)>\mathcal{B}\left(B\rightarrow K^{(*)}\nu\nu\right)^{\mathrm{SM}}$. Thus, the orange region is not visible because it lies below the hatched pink band. In the right panel of Fig.~\ref{fig:BR_BtoK_vs_BR_BtoKst} we show the impact of such operators on the fraction of longitudinally polarized $K^*$, $F_L$~\cite{Buras:2014fpa}, which can also be measured at Belle II.~\footnote{We refer the reader to Appendix~\ref{sec:diff_dec_rates} for a definition of $F_L$ in terms of the $q^2$-dependent functions entering the differential branching fraction.} In particular we study the correlation between $\mathcal{B}\left(B\rightarrow K\nu\nu\right)$ and the ratio $\mathcal{R}_{F_L}=F_L/F_{L}^{\mathrm{SM}}$, with  $F_L^{\mathrm{SM}}=0.49(4)$~\cite{Becirevic:2023aov}. We show, as solid bands, the $1\sigma$ regions for which the Belle bound $\mathcal{B}\left(B\rightarrow K^*\nu\nu\right)<2.7\times 10^{-5}$ is satisfied. The (dash-) dotted lines allow to understand the behavior of $\mathcal{R}_{F_L}$ in terms of $\mathcal{B}\left(B\rightarrow K\nu\nu\right)$ in the whole parameter space, but correspond to regions where the branching fraction of $B\rightarrow K^*\nu\nu$ is too large. They are thus experimentally excluded. 

For the operators which can accommodate $\mathcal{B}\left(B\rightarrow K\nu\nu\right)^{\mathrm{exp}}$ at $2\sigma$ in this mass range, we find the following allowed values for the WCs (computed at the scale $\mu=1$~TeV). First, for the operators involving only SM neutrinos, we find:
\begin{equation}
    \begin{split}
            \centerdot \;&\left[\mathcal{C}_{ld}\right]_{\tau\tau}/\Lambda^2\in\left[-3.3,-1.1\right]\times 10^{-2}\,\mathrm{TeV}^{-2}\,,\\
        &\mathcal{B}\left(B\rightarrow K^*\nu\nu\right)\in\left[8,27\right] \times 10^{-6}\,, \\
        &\mathcal{R}_{F_L}\in\left[0.73,0.76\right]\,.\\
        \centerdot \;&\left[\mathcal{C}^{(1-3)}_{lq}\right]_{\tau\tau}/\Lambda^2\in\left[-1.7,-1.1\right]\cup\left[3.2,3.7\right]\times 10^{-2}\,\mathrm{TeV}^{-2}\,,\\
        &\mathcal{B}\left(B\rightarrow K^*\nu\nu\right)\in\left[19,27\right]\times 10^{-6}\,, \\
        &\mathcal{R}_{F_L} = 1\,.\\
    \end{split}
    \label{eq:ranges_m0_nuL}
\end{equation}
Next, for operators involving the RH neutrino:
\begin{equation}
    \begin{split}
            \centerdot \;&\Big|\left[\mathcal{C}^S_{lNqd}\right]_{\tau 23}\Big|/\Lambda^2\in\left[0.7,1.7\right]\times 10^{-2}\,\mathrm{TeV}^{-2}\,, \\
            &\mathcal{B}\left(B\rightarrow K^*\nu\nu\right)\in\left[11,21\right]\times 10^{-6}\,, \\ &\mathcal{R}_{F_L}\in\left[1.19,1.61\right]\,.\\
        \centerdot \;&\Big|\left[\mathcal{C}_{Nq}\right]\Big|/\Lambda^2\in\left[1.9,2.5\right]\times 10^{-2}\,\mathrm{TeV}^{-2}\,,\\
        &\mathcal{B}\left(B\rightarrow K^*\nu\nu\right)\in\left[19,27\right]\times 10^{-6}\,, \\
        &\mathcal{R}_{F_L} = 1\,.\\
    \end{split}
    \label{eq:ranges_m0_NR}
\end{equation}

Note that for the scalar operator with exchanged quark indexes, $2\leftrightarrow 3$, the same results as those from the first line in Eq.~(\ref{eq:ranges_m0_NR}) would be obtained. Similarly, the effect of $\mathcal{O}_{Nd}$ is degenerate with that of $\mathcal{O}_{Nq}$ at this level.

Remarkably, operators with SM neutrinos accommodating $\mathcal{B}\left(B\rightarrow K\nu\nu\right)^{\mathrm{exp}}$ at $1\sigma$, such as $\mathcal{O}_{ld}$, would decrease $F_L$ with respect to its SM value, whereas those with RH neutrinos would enhance it, providing a handle to distinguish these solutions. We will show next that this feature holds also for non-negligible heavy neutrino masses.
\item $m_{n_4}=2$~GeV
\begin{figure*}
\centering
    \includegraphics[width=0.45\textwidth]{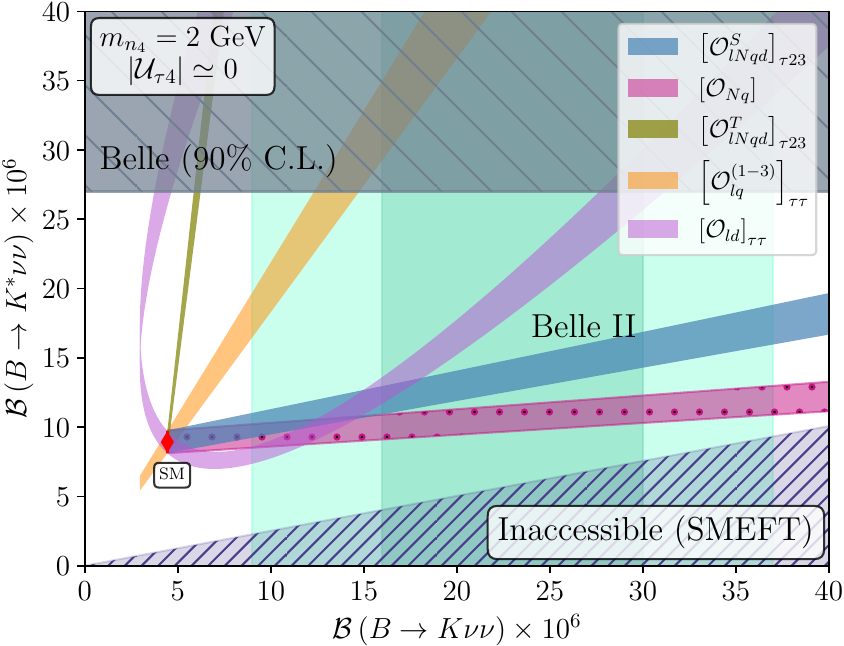}
    \includegraphics[width=0.45\textwidth]{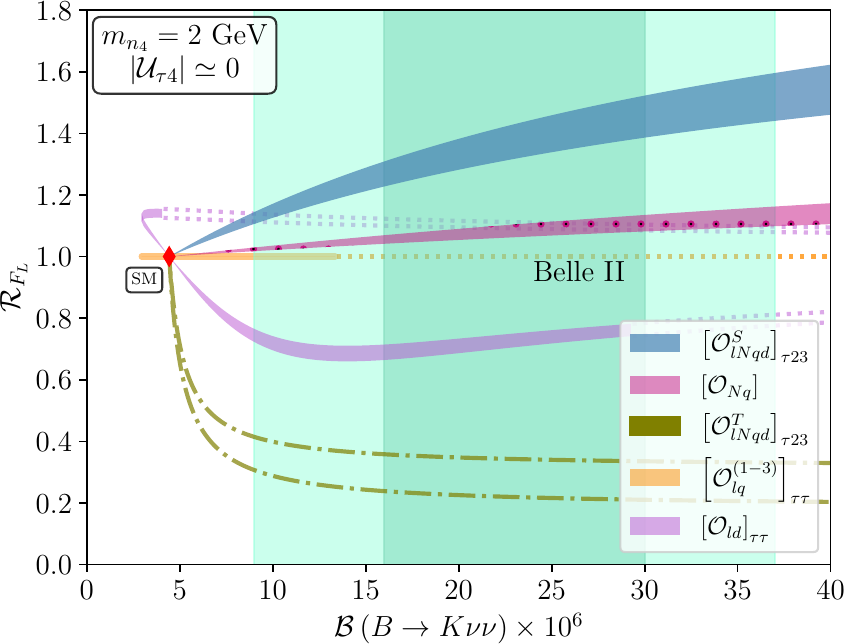}
    \caption{\small \sl Correlation between $\mathcal{B}\left(B\rightarrow K\nu\nu\right)$ and $\mathcal{B}\left(B\rightarrow K^*\nu\nu\right)$ (left panel) and $\mathcal{R}_{F_L}$ (right panel), as in Fig.~\ref{fig:BR_BtoK_vs_BR_BtoKst}, but setting now the mass of the heavy neutrino to $m_{n_4}=2$~GeV.}
    \label{fig:BR_BtoK_vs_FL}
\end{figure*}

As previously mentioned, increasing the mass of the heavy neutrino tends to suppress the available phase space, and thus the $B\rightarrow K^{(*)}\nu\nu$ branching fraction. This is particularly pronounced in the case of the vector operator, $\mathcal{O}_{Nq(d)}$, as shown in the left panel of Fig.~\ref{fig:BR_BtoK_vs_FL} where we set $m_{n_4}=2$~GeV. This mass is very close to the kinematical limit for the $B\rightarrow K^*\nu\nu$ decay, as the vector operator couples directly to a pair of RH neutrinos. Given that the $N_R$ is mostly aligned with the heavy neutrino, the contribution from  this operator is suppressed whenever $m_{n_4}\sim (m_B-m_{K^*})/2$.~\footnote{Neglecting the active-heavy mixing, it is exactly aligned with the heavy neutrino, meaning that $N_R=P_R n_4$.} Instead, for the scalar and tensor operators in Eq.~(\ref{eq:Op_nuSMEFT}), the phase space suppression in $B\rightarrow K^*\nu\nu$ becomes relevant only when $m_{n_4}\sim m_B-m_{K^*}$. In fact, it is only through such a phase space suppression that the tensor operator can accommodate the Belle II result on $B\rightarrow K\nu\nu$ without predicting values of $\mathcal{B}\left(B\rightarrow K^*\nu\nu\right)$ above the experimental bound. 

We show in the right panel of Fig.~\ref{fig:BR_BtoK_vs_FL} the enhancement produced on $F_L$ by the vector operator with RH neutrinos, $\mathcal{O}_{Nq(d)}$, when having a non-negligible mass. This effect is, however, not as pronounced as with the scalar operator. 

Due to the phase space suppression, $\mathcal{O}_{Nq}$ ($\mathcal{O}_{Nd}$)  can now also accommodate $\mathcal{B}\left(B\rightarrow K\nu\nu\right)^{\mathrm{exp}}$ at $1\sigma$. We quote next the allowed ranges for the WCs for $m_{n_4}=2$~GeV:~\footnote{We explicitly write again the ranges for the scalar operator as it is affected by the phase space suppression, even if to a lesser degree than the vector operator. On the contrary, we note that for the SMEFT operator $\mathcal{O}_{ld}\,(\mathcal{O}_{lq}^{(1-3)})$ the ranges are those from Eq.~\ref{eq:ranges_m0_nuL}.} 
\begin{equation}
    \begin{split}
            \centerdot \;&\Big|\left[\mathcal{C}^S_{lNqd}\right]_{\tau 23}\Big|/\Lambda^2\in\left[1.0,2.5\right]\times 10^{-2}\,\mathrm{TeV}^{-2}\,, \\
            &\mathcal{B}\left(B\rightarrow K^*\nu\nu\right)\in\left[10,17\right]\times 10^{-6}\,, \\ &\mathcal{R}_{F_L}\in\left[1.14,1.51\right]\,.\\
            \centerdot \;&\Big|\left[\mathcal{C}_{Nq}\right]\Big|/\Lambda^2\in\left[4.5,11.2\right]\times 10^{-2}\,\mathrm{TeV}^{-2}\,,\\
        &\mathcal{B}\left(B\rightarrow K^*\nu\nu\right)\in\left[9,12\right]\times 10^{-6}\,, \\
        &\mathcal{R}_{F_L}\in\left[1.02,1.13\right]\,.\\
    \end{split}
\end{equation}
\end{itemize}

In conclusion, for negligible $m_{n_4}$, we find that the scalar operator $\mathcal{O}_{lNqd}^S$ and the vector one, $\mathcal{O}_{ld}$, represent the best possibilities of explaining the Belle II result at the $1\sigma$ level. The other vector operators can, nonetheless, accommodate the experimental measurement at $2\sigma$, while the tensor one is ruled out experimentally. 

For larger masses, the phase space suppression allows to explain $\mathcal{B}\left(B\rightarrow K\nu\nu\right)^{\mathrm{exp}}$, without enhancing $\mathcal{B}\left(B\rightarrow K^*\nu\nu\right)$ above its experimental bound, using operators involving $N_R$. This is relevant for $m_{n_4}\sim(m_B-m_{K^*})/2$ for vector operators and $m_{n_4}\sim m_B-m_{K^*}$ for the tensor one.

Interestingly, when studying the effect of such operators on $F_L$, we find opposite behaviours for different types of operators accommodating the Belle II result at the $1\sigma$ level. Those involving only SM neutrinos, namely $\mathcal{O}_{ld}$, tend to suppress $F_L$, such that we have $\mathcal{R}_{F_L}\leq 1$, as previously found in Ref.~\cite{Allwicher:2023xba}. Instead, operators involving $N_R$ enhance it, so that $\mathcal{R}_{F_L}\geq 1$. Thus, this measurement could allow to tell apart between contributions from $N_R$ and those from SM neutrinos, assuming a single operator is responsible for $\mathcal{B}\left(B\rightarrow K\nu\nu\right)^{\mathrm{exp}}$.

\subsubsection{Effect of neutrino mixing}
We focus now on the impact neutrino mixing can have on the phenomenology, setting the mixing between the $\tau$ neutrino and the heavy one to $|\mathcal{U}_{\tau 4}|=10^{-2}$.~\footnote{This is just a representative value. For heavy neutrino masses lying around the GeV-scale, one finds constraints at the level of $|\mathcal{U}_{\tau 4}|^2\sim 10^{-4}-10^{-5}$~\cite{Fernandez-Martinez:2023phj} and in agreement with our choice. A somewhat smaller mixing would translate into a larger value for the relevant WC while generating the same effect on the observables.} The consequence is that now contributions from operators involving the RH neutrino (see Eq.~\ref{eq:Op_nuSMEFT}) will also have an impact on the $b\rightarrow s\nu\nu$ decay into light neutrinos. This will in turn allow for the interference between contributions from $\mathcal{O}_{Nq}$ ($\mathcal{O}_{Nd}$) and the SM. The most important difference with respect to $|\mathcal{U}_{\tau 4}|=0$ arises close to the kinematical limit for the $B\rightarrow K^{(*)}\nu\nu$ decay into one heavy neutrino, i.e. $m_{n_4}\gtrsim m_B-m_K$. 

\begin{itemize}
    \item $m_B-m_K\lesssim m_{n_4} \lesssim m_{B_s}$
    
In this mass range, the bound $\mathcal{B}\left(B_s\rightarrow \nu\nu\right)<5.9\times 10^{-4}$ at 90\% C. L.~\cite{Alonso-Alvarez:2023mgc} becomes the most relevant one. This constraint allows to exclude explanations of $\mathcal{B}\left(B\rightarrow K\nu\nu\right)^{\mathrm{exp}}$ with scalar and vector operators involving $N_R$, as shown in Fig.~\ref{fig:BR_BtoK_vs_Bs}. This pushes the allowed range for the heavy neutrino mass to be either $m_{n_4}>m_{B_s}$ or $m_{n_4}\lesssim 4$~GeV for the operators $\mathcal{O}_{lNqd}^S$ and $\mathcal{O}_{Nq}$ ($\mathcal{O}_{Nd}$). 
\begin{figure*}
\centering
    \includegraphics[width=0.45\textwidth]{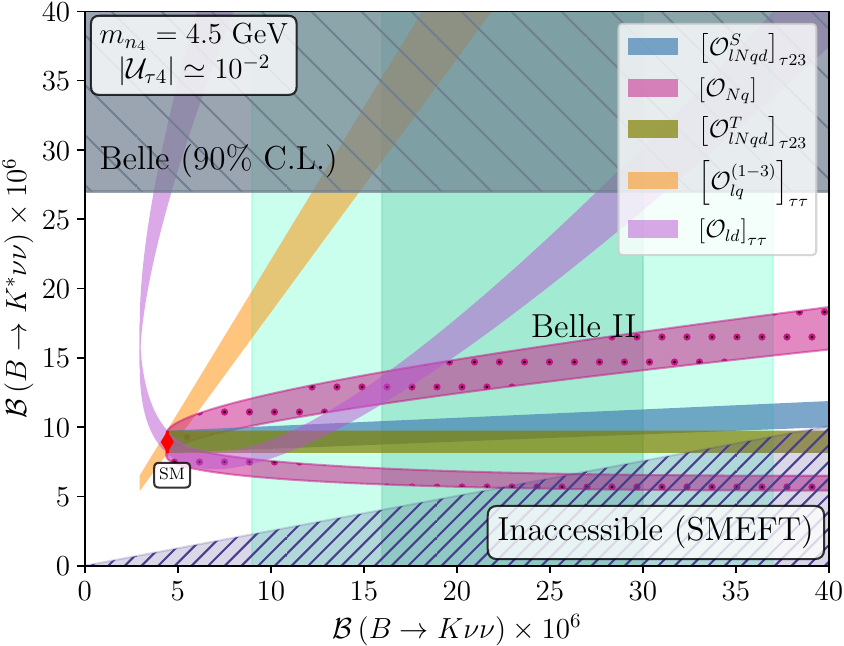}
    \includegraphics[width=0.45\textwidth]{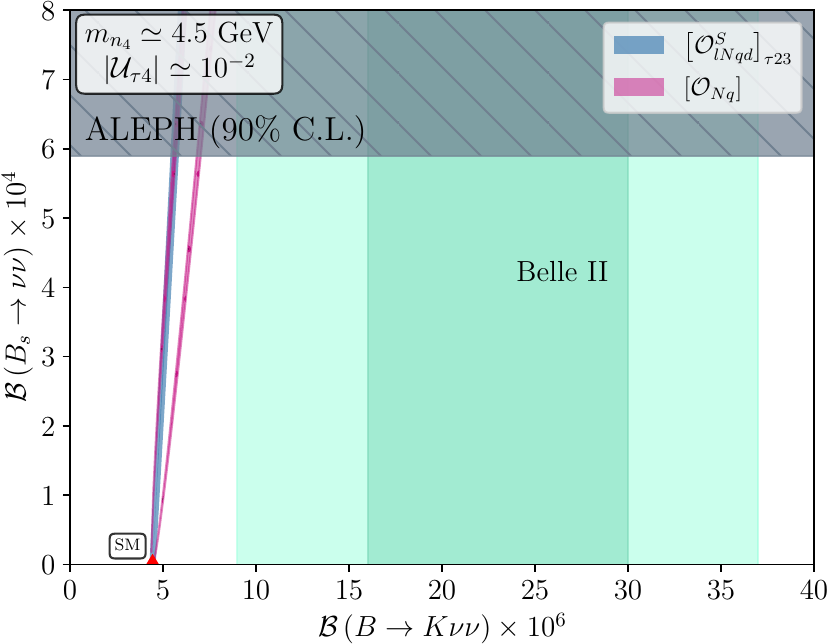}
    \caption{\small \sl Correlation between different observables when including the effect of one single NP operator at a time for $m_{n_4}=4.5$~GeV and non-zero mixing, $|\mathcal{U}_{\tau 4}|\sim 10^{-2}$. The red point corresponds to the SM expectation and the (light) green regions to the Belle II result at $1(2)\sigma$. The left panel shows the relation between $\mathcal{B}\left(B\rightarrow K\nu\nu\right)$ and $\mathcal{B}\left(B\rightarrow K^*\nu\nu\right)$ while in the right one we show the correlation with $\mathcal{B}\left(B_s\rightarrow \nu\nu\right)$ for the relevant operators.}
    \label{fig:BR_BtoK_vs_Bs}
\end{figure*}

Instead, the tensor operator can accommodate the experimental result for $m_{n_4}\simeq m_B-m_K$, as shown in the left panel in Fig.~\ref{fig:BR_BtoK_vs_Bs}. We find the following range for the tensor WC:~\footnote{This range applies for the tensor operator independently of $|\mathcal{U}_{\tau 4}|$.}
\begin{equation}
\begin{split}
    \centerdot \;&\left[\mathcal{C}^T_{lNqd}\right]_{\tau 23}/\Lambda^2\in \left[0.71,\,1.58\right]\,\mathrm{TeV}^{-2}\,,\\
    &\mathcal{B}\left(B\rightarrow K^*\nu\nu\right)=\mathcal{B}\left(B\rightarrow K^*\nu\nu\right)^{\mathrm{SM}}\,,\\
    &\mathcal{R}_{F_L}=1\,.
    \end{split}
    \label{eq:range_tensor}
\end{equation}
However, as we will see, this scenario (when involving the second generation quark doublet) is excluded by the LHC constraints we discuss in Section~\ref{sec:LHC}.

\item $m_{n_4}>m_{B_s}$

Whenever $n_4$ cannot be produced in $B_s\rightarrow \nu\nu$, the effect of $\mathcal{O}_{Nd}$ ($\mathcal{O}_{Nq}$) becomes degenerate with that of the SMEFT operator $\mathcal{O}_{ld}$ ($\mathcal{O}_{lq}^{(1-3)}$), while avoiding any other constraint. Indeed, the sum over neutrinos in $\delta\mathcal{B}_{K^{(*)}}$ from Eq.~(\ref{eq:deltaBK}) would only run through the three light neutrinos, recovering the equivalent results in SMEFT~\cite{Allwicher:2023xba,Marzocca:2024hua}. 

Note, however, that the contribution from operators involving a RH neutrino would be mixing suppressed. Thus, in order to accommodate the Belle II result with $\mathcal{O}_{Nd}$, the size of the WC needs to be, in comparison to the corresponding SMEFT contribution, $\mathcal{C}_{Nd}= |\mathcal{U}_{\tau 4}|^{-2}\mathcal{C}_{ld}$. Interpreting once again the corresponding WC, $\mathcal{C}_{Nd}/\Lambda^2$, in terms of a tree-level contribution from a heavy NP mediator, we find in this case that the maximum cutoff scale is $\Lambda\lesssim 240$~GeV. Such a low scale clearly signals the breakdown of the EFT, and thus we deem this possibility not viable. We conclude then that the vector operator $\mathcal{O}_{Nd}$ ($\mathcal{O}_{Nq}$) can only accommodate the experimental result for $m_{n_4}<4$~GeV, assuming NP contributes to $b\rightarrow s\nu\nu$ at tree-level.

Regarding the scalar operator, its contribution is equivalent to the case we considered in Section~\ref{sec:NoNuMix}, summarized in Eq.~(\ref{eq:ranges_m0_NR}). Given, however, the mixing suppression, the size of the WC needs to be rescaled, exchanging $\mathcal{C}^S_{lNqd}\rightarrow |\mathcal{U}_{\tau 4}|\mathcal{C}^S_{lNqd}$ in Eq.~(\ref{eq:ranges_m0_NR}). Given current bounds on the mixing, we will show next that $\left[\mathcal{O}_{lNqd}^S\right]_{\tau 23}$ is excluded by high-$p_T$ constraints. 
\end{itemize}

\subsection{Implications for $b\rightarrow c\ell_{\alpha}\nu$}
As previously discussed, given the $SU(2)_L$ gauge symmetry, the same WCs generating $b\rightarrow s\nu\nu$ at low energies can also contribute to processes involving charged leptons. 

In the case of the SMEFT operators in Eq.~(\ref{eq:Op_SMEFT}), they all simultaneously contribute to $b\rightarrow s\nu\nu$ and $b\rightarrow s\ell_{\alpha}\ell_{\beta}$. Stringent constraints exist for the latter process when the leptons are electrons or muons~\cite{ParticleDataGroup:2022pth}. In such a situation, the only solution is to introduce the NP effects in the third generation of leptons only~\cite{Bause:2023mfe, Allwicher:2023xba}. It was found that, even if $\mathcal{O}^{(3)}_{lq}$ can contribute to $b\rightarrow c\tau \nu$, its potential to also explain the current experimental deviation from the SM prediction of the ratio $R_{D^{(*)}}=\mathcal{B}\left(B\rightarrow D^{(*)}\tau\bar{\nu}\right)/\mathcal{B}\left(B\rightarrow D^{(*)}\ell\bar{\nu}\right)$ with $\ell=e,\mu$ is very limited. Only a marginal enhancement of $R_{D^{(*)}}$ can be expected while agreeing with the Belle II result. 

Regarding the operators involving RH neutrinos in Eq.~(\ref{eq:Op_nuSMEFT}), only the scalar and tensor operators with a second generation quark doublet, namely $\left[\mathcal{O}^{S(T)}_{lNqd}\right]_{\tau 23}$, can contribute to $R_{D^{(*)}}$. However, none of them can simultaneously accommodate the experimental results on $R_{D^{(*)}}$ and $\mathcal{B}\left(B\rightarrow K\nu\nu\right)^{\mathrm{exp}}$. As it is clear from Eq.~(\ref{eq:rel_btoc_btos}), only very large NP effects in $b\rightarrow s\nu\nu$, predicting $\mathcal{B}\left(B\rightarrow K\nu\nu\right)>3\times 10^{-4}$, can introduce a significant deviation from the SM expectation in $B\rightarrow D^{(*)}\tau\nu$, given the factor $V_{cs}\alpha_{\mathrm{em}}\lambda_t/(\pi V_{cb})\sim 10^{-3}$ relating the WCs entering the two processes. However, this is obviously excluded experimentally. Moreover, the tensor operator cannot contribute to $R_{D^{(*)}}$ in the region where it accommodates $\mathcal{B}\left(B\rightarrow K\nu\nu\right)^{\mathrm{exp}}$, as one needs $m_{n_4}\gtrsim m_B-m_{K^*}>m_B-m_{D^{(*)}}-m_{\tau}$.

\subsection{LHC bounds}
\label{sec:LHC}
We now turn to consider collider constrains on these same WCs through the crossed channel $bc\rightarrow \ell_{\alpha}\nu$. In particular, we can profit from the effect of such EFTs on the high-$p_T$ tails of Drell-Yan processes with a charged lepton and missing energy in the final state~\cite{Farina:2016rws}.~\footnote{One can also use constrains from channels with large missing energy and an energetic jet in the final state~\cite{Hiller:2024vtr}.} LHC has performed such searches~\cite{ATLAS:2019lsy,ATLAS:2024tzc}, $pp\rightarrow \ell_{\alpha}\nu$ with $\alpha=e,\mu$ or $\tau$, which have been fully implemented in {\tt HighPT}~\cite{Allwicher:2022mcg,Allwicher:2022gkm}, alongside the relevant $d=6$ SMEFT operators. Thus, {\tt HighPT} generates the likelihoods for different NP scenarios, allowing to constrain the SMEFT and $\nu$SMEFT operators generating $b\rightarrow c\ell_{\alpha}\nu$. It is important to note here that, for the scalar and tensor $\nu$SMEFT operators from Eq.~(\ref{eq:Op_nuSMEFT}), we recast the limits obtained with {\tt HighPT} in terms of the following SMEFT ones (which do not interfere with the SM contribution to $pp\rightarrow\ell_{\alpha}\nu$): 
\begin{equation}
    \begin{split}           \left[\mathcal{O}^{(1)}_{lequ}\right]_{\beta\alpha 32}&=\left(\bar{L}_{\beta}\ell_{\alpha R}\right)\epsilon\left(\bar{Q}_{3}u_{2R}\right),\\
    \left[\mathcal{O}^{(3)}_{lequ}\right]_{\beta\alpha 32}&=\left(\bar{L}_{\beta}\sigma_{\mu\nu}\ell_{\alpha R}\right)\epsilon\left(\bar{Q}_{3}\sigma^{\mu\nu}u_{2R}\right).
    \end{split}
\end{equation}
Using the mono-lepton searches at LHC, which do not tag the final neutrino flavor, we can use the constraints on the SMEFT WCs to find a bound on the $\nu$SMEFT scalar and tensor WC involving a charged lepton, $\ell_{\alpha}$, as:
\begin{equation}
\begin{split}
    |V_{cs}|^2\left|\left[\mathcal{C}^S_{lNqd}\right]_{\alpha 23}\right|^2&\rightarrow \sum_{\beta=e,\mu,\tau}\left[\mathcal{C}^{(1)}_{lequ}\right]_{\beta\alpha}^*\left[\mathcal{C}^{(1)}_{lequ}\right]_{\beta\alpha},\\
    |V_{cs}|^2\left|\left[\mathcal{C}^T_{lNqd}\right]_{\alpha 23}\right|^2&\rightarrow \sum_{\beta=e,\mu,\tau}\left[\mathcal{C}^{(3)}_{lequ}\right]_{\beta\alpha}^*\left[\mathcal{C}^{(3)}_{lequ}\right]_{\beta\alpha},
\end{split}
\end{equation}
where we neglect the heavy neutrino mass in the final state, thus making the contribution from $\mathcal{O}^{(1)}_{lequ}$ equivalent to that of the $\mathcal{O}^S_{lNqd}$ operator involving the second generation $SU(2)_L$ quark doublet. Note that for the scalar and tensor $\nu$SMEFT operators involving the third generation quark doublets, we would in turn be sensitive to the semileptonic decay of the top quark, $t\rightarrow s\ell_{\alpha} \nu$. This, however, does not contribute to $pp\rightarrow \ell_{\alpha}\nu$ at leading order. Moreover, the experimental precision on $\mathcal{B}\left(t\rightarrow s\tau \nu\right)\simeq \left(5\pm4\right)\times 10^{-3}$ is weak~\cite{ParticleDataGroup:2022pth}, not leading to useful constraints.~\footnote{We estimate it by combining $\mathcal{B}\left(t\rightarrow Wb\right)$ and $\mathcal{B}\left(W\rightarrow \tau \nu\right)$ with $\mathcal{B}\left(t\rightarrow Wb\right)/(\mathcal{B}\left(t\rightarrow Wb\right)+\mathcal{B}\left(t\rightarrow Ws\right))=0.957\pm 0.034$.}

Using {\tt HighPT}, we find then the following bounds at $90\%$~C.L. on the scalar and tensor WC:
\begin{equation}
\begin{split}
    \left|\left[\mathcal{C}^S_{lNqd}\right]_{\tau 23}\right|/\Lambda^2\lesssim 0.41\,\mathrm{TeV}^{-2},\\
    \left|\left[\mathcal{C}^T_{lNqd}\right]_{\tau 23}\right|/\Lambda^2\lesssim 0.25\,\mathrm{TeV}^{-2}.
\end{split}
\label{eq:highPT}
\end{equation}
We quote here the results for those WCs involving the $\tau$-lepton, but note that those involving an electron or a muon in the final state are quantitatively very similar. 

These bounds can now be used to constrain the effect of $\left[\mathcal{O}^{S(T)}_{lNqd}\right]_{\tau 23}$  on $B\rightarrow K\nu\nu$. They are relevant for the scalar operator for $m_{n_4}>m_{B_s}$, as, due to the mixing suppression, the size of the WC becomes very large. Indeed, for the region accommodating the Belle II result, we find 
\begin{equation}
    \Big|\left[\mathcal{C}^S_{lNqd}\right]_{\tau 23}\Big|/\Lambda^2\gtrsim 0.7\,\mathrm{TeV}^{-2}\,,
\end{equation}
excluded by the bound in Eq.~(\ref{eq:highPT}).~\footnote{Note here that we used a particular value of the neutrino mixing, $|\mathcal{U}_{\tau 4}|=10^{-2}$.} They also provide the main constraint on the tensor operator for large $m_{n_4}$. As shown in Eq.~(\ref{eq:range_tensor}), the tensor operator accommodates $\mathcal{B}\left(B\rightarrow K\nu\nu\right)^{\mathrm{exp}}$ for values of the WC larger than those allowed by the high-$p_T$ analysis. Given that this operator only works when $m_{n_4}\sim m_B-m_{K^*}$, we conclude that this explanation of the Belle II result is experimentally excluded by LHC data.

\section{Summary}
\label{sec:conclusions}
In this letter we presented the study of the contribution from operators in the $\nu$SMEFT, both involving only SM fields as well as one RH neutrino, to the rare semileptonic $b\rightarrow s\nu\nu$ decays. The presence of the RH neutrino in the EFT description introduces the possibility to have scalar and tensor contributions to this transition, on top of the usual vector operators. Besides studying the impact of the heavy neutrino mass on $B\rightarrow K^{(*)}\nu\nu$, we also studied the effect of its mixing with the SM neutrinos. Examining the impact of such NP on the correlations between the low-energy observables, namely $\mathcal{B}\left(B\rightarrow K^{(*)}\nu\nu\right)$ and $F_L$, the fraction of longitudinally polarized $K^*$ in the final state, we were able to identify the best solutions accommodating the recent Belle II measurement. 

First, we studied the impact of these operators in the absence of neutrino mixing. We find that for a light RH neutrino, the only NP operators providing a plausible explanation to the Belle II result at $1\sigma$ are the scalar operator $\mathcal{O}^S_{lNqd}$ involving a RH neutrino, and the vector operator $\mathcal{O}_{ld}$ with SM neutrinos. This is consistent with previous results from Refs.~\cite{Bause:2023mfe,Allwicher:2023xba,Felkl:2023ayn}. At the $2\sigma$ level, other vector operators are able to accommodate the Belle II measurement. The tensor operator is ruled out, as it completely pushes $\mathcal{B}\left(B\rightarrow K^*\nu\nu\right)$ above the experimental bound $\mathcal{B}\left(B\rightarrow K^*\nu\nu\right)<2.7\times 10^{-5}$. 

However, if the mass of the RH neutrino is non-negligible, then even the tensor operator can accommodate $\mathcal{B}\left(B\rightarrow K\nu\nu\right)^{\mathrm{exp}}$ at the $1\sigma$ level. This is due to the phase space suppression in the $B\rightarrow K^*\nu\nu$ decay, which becomes relevant for the  tensor and scalar operators, $\mathcal{O}^{T(S)}_{lNqd}$ when 
$m_{n_4}\gtrsim m_B-m_{K^*}\simeq 4.5$~GeV. In the case of the vector operator $\mathcal{O}_{Nd}$ ($\mathcal{O}_{Nq}$), the suppression is significant for $m_{n_4}\gtrsim(m_B-m_{K^*})/2\simeq 2$~GeV.

Along with what was noted in Refs.~\cite{Bause:2023mfe,Allwicher:2023xba,Felkl:2023ayn} we also find correlations between $b\rightarrow s\nu\nu$ and $b\rightarrow c\ell_{\alpha}\nu$ in $\nu$SMEFT. This is why we considered the impact of operators involving the RH neutrino on $B\rightarrow D^{(*)}\tau\nu$ and LHC searches involving high-$p_T$. However, we find that the operators explaining the Belle II result cannot contribute significantly to $\mathcal{B}\left(B\rightarrow D^{(*)}\tau\nu\right)$. Therefore it is not possible to simultaneously explain $\mathcal{B}\left(B\rightarrow K\nu\nu\right)^{\mathrm{exp}}$ and the $R_{D^{(*)}}$ anomalies. On the other hand, the analysis of high-$p_T$ tails of Drell-Yan processes at LHC allows us to obtain rather strong constraints on the coupling to the tensor operator. From that analysis we conclude that the Belle II result cannot be described with the tensor operator involving the second generation $SU(2)_L$ quark doublet. 

We also studied $\mathcal{R}_{F_L}=F_L/F_L^{\mathrm{SM}}$ and found that operators involving SM neutrinos cannot enhance $F_L$, i.e. $\mathcal{R}_{F_L}\leq 1$. Instead, operators with RH neutrinos explaining Belle II always translate into $\mathcal{R}_{F_L}\geq 1$. Thus, the measurement of this observable could allow to tell apart contributions involving RH neutrinos from those with only SM neutrinos. Moreover, measuring the differential decay rates could help us disentangling among various Lorentz structures of the operators. This observation is in line with Refs.~\cite{Altmannshofer:2023hkn,Felkl:2023ayn,Bolton:2024egx}. 

Finally, we turned then to study the impact of the mixing between the SM neutrinos and the heavy one, for which experimental constraints show that $|\mathcal{U}_{\tau 4}|\lesssim 10^{-2}$. Despite its smallness, it can lead to interesting phenomenology. In this case, even if the RH neutrino is not produced in the $B\rightarrow K \nu\nu$ decay, i.e. $m_{n_4}>m_B-m_K$, it can make an impact on $b\rightarrow s\nu\nu$ through its mixing with the SM neutrinos. We find that, in order to avoid the bound from LEP on $\mathcal{B}\left(B_s\rightarrow \nu\nu\right)$ with scalar and vector operators, the mass of the heavy neutrino needs to be $m_{n_4}\lesssim 4$~GeV (regardless of mixing) or $m_{n_4}>m_{B_s}$. In the latter case, only the scalar operator involving the third generation quark doublet can consistently describe the Belle II result, predicting $\mathcal{B}\left(B_s\rightarrow\nu\nu\right)\gtrsim 10^{-4}$.

In principle, one can find an useful observable from the angular distribution of $B\rightarrow K\nu\nu$, whose size is proportional to the neutrino masses in the final state for vector and scalar NP operators. However, it is not experimentally accessible given that the neutrinos escape the detector. We note as well that the same NP affecting $B\rightarrow K^{(*)}\nu\nu$ can be probed studying complementary baryon decays such as $\Lambda_b\rightarrow \Lambda_c \ell \nu$~\cite{Becirevic:2022bev} and potentially also $\Lambda_b\rightarrow \Lambda\nu\nu$~\cite{Amhis:2023mpj}.

\section*{Acknowledgments} 
The authors warmly thank Olcyr Sumensari and Damir Becirevic for innumerable discussions and encouragement, as well as for reading the first version of the manuscript. We also acknowledge very helpful discussions with Adam Falkowski. This project has received funding /support from the European Union’s Horizon 2020 research and innovation programe under the Marie Skłodowska-Curie grant agreement No 860881-HIDDeN and No 101086085-ASYMMETRY, as well as from SPRINT FAPESP/CNRS No. 2023/00643-0. L.P.~S.~Leal is fully financially supported by FAPESP under Contracts No. 2021/02283-6 and No. 2023/12330-7. L.P.~S.~Leal also would like to thank the hospitality of 
the IJCLab Theory Group. 

\appendix
\section{Neutrino masses}
\label{app:Neutrino_masses}
We devote this appendix to a brief summary of the generation of neutrino masses in the presence of RH neutrinos, and the relation between the mass and flavor bases. Adding $n_R$ RH neutrinos to the SM particle content, we can write the following lagrangian for the neutrino sector
\begin{equation}
	\begin{split}
		\mathcal{L}\supset -\bar{L}Y_{\nu}\tilde{H}N_R-\frac{1}{2}\bar{N}_R^cMN_R+h.c.
	\end{split}
	\label{eq:lag_nuMass}
\end{equation}
where $\tilde{H}=i\tau_2 H^*$ is the Higgs field and $Y_{\nu}$ corresponds to the Yukawa matrix for the neutrinos. $M$ is the Majorana mass matrix for the $n_R$ RH neutrinos. Note that this Majorana mass is not related to the Higgs mechanism and thus its scale is completely free. After spontaneous symmetry breaking (SSB) we find the following mass term for the neutrinos
\begin{equation}
	\mathcal{L}\supset -\frac{1}{2}\begin{pmatrix} \bar{\nu}_L & \bar{N}_R^c\end{pmatrix}
	\begin{pmatrix} 0 & m_D\\
	m_D^T & M\end{pmatrix} 
	\begin{pmatrix} \nu_L^c\\ 
	N_R\end{pmatrix} + h.c.\,,
 \label{eq:mass_matrix}
\end{equation}
where $m_D\equiv v Y_{\nu}/\sqrt{2}$. The symmetric mass matrix in Eq.~(\ref{eq:mass_matrix}) can then be diagonalized through a unitary rotation, $\mathcal{U}$, such that one finds the following relation between the flavor and the mass eigenstates
\begin{equation}
	\begin{pmatrix} \nu_{\alpha L}\\
	N_{sR}^c\end{pmatrix}
	= \sum_{i=1}^{3+n_R}\begin{pmatrix} \mathcal{U}_{\alpha i}\\
	\mathcal{U}_{s i}\end{pmatrix}
	P_L  n_i \,,
\end{equation}
where we write the full neutrino mixing matrix, $\mathcal{U}$, in two blocks. The upper one corresponding to the mixing between the active SM neutrinos, $\nu_{\alpha L}$, and the mass eigenstates given by $\mathcal{U}_{\alpha i}$ which is a $3\times (3+n_R)$ matrix. The lower one describes the mixing between the RH neutrinos and the massive ones, $\mathcal{U}_{s i}$, whose dimensions would be $n_R\times (3+n_R)$. In the seesaw limit, namely $m_D\ll M$, we find three light states whose mass is given by $m_{\mathrm{light}}\sim -m_DM^{-1}m_D^T$ and $n_R$ heavy neutrino states with masses $\mathcal{O}\left(M\right)$.

Depending on the mass scale of the heavy neutrinos, constrains on the size of the active-heavy mixing, $\mathcal{U}_{\alpha i}$ with $i=4,...,3+n_R$, can be placed. In particular, for the mixing with the $\tau$-sector, and masses $m_{n_4}\sim 2$~GeV constraints from DELPHI exist~\cite{DELPHI:1996qcc}, such as $|\mathcal{U}_{\tau 4}|^2\lesssim 10^{-4}$~\cite{Fernandez-Martinez:2023phj}. For larger masses, $m_{n_4}\sim 10$~GeV, one instead finds slightly stronger bounds, at the level of $|\mathcal{U}_{\tau 4}|^2\lesssim 10^{-5}$.

\section{Form factors}
\label{sec:Form_factors}
In the following we will summarize the parametrization of the hadronic amplitudes in terms of form factors both for $B\rightarrow P$, with $P=K,\,D$ and $B\rightarrow V$, where $V=K^*,\,D^*$.
\subsection{$B\rightarrow P$}
The relevant hadronic matrix elements for the $B\rightarrow P$ transition can be written as
\begin{equation}
	\begin{split}
		\big\langle \bar{P}(k)|\bar{q}_i\gamma^{\mu}b|\bar{B}(p)\big\rangle=&\left[\left(p+k\right)^{\mu}-\frac{m_B^2-m_P^2}{q^2}q^{\mu}\right]f_+(q^2)\\
		+&\frac{m_B^2-m_P^2}{q^2}q^{\mu}f_0(q^2)\,,\\
		\big\langle \bar{P}(k)|\bar{q}_i\sigma^{\mu\nu}b|\bar{B}(p)\big\rangle=&-i\left[p^{\mu}k^{\nu}-p^{\nu}k^{\mu}\right]\frac{2f_T(q^2)}{m_B+m_P}\,,
	\end{split}
	\label{eq:Vector_FF_BtoK}
\end{equation}	
where $q^2=(p-k)^2$ is the momenta carried by the di-neutrino system. For the $B\rightarrow K$ transition we have $P=K$ and $q_i=s$, the strange quark, in Eq.~(\ref{eq:Vector_FF_BtoK}), while $P=D$ and $q_i=c$ (charm quark) for $B\rightarrow D$. The vector and scalar form factors satisfy $f_+=f_0$ at $q^2=0$. 

We take the form factors for $B\rightarrow K$ from Ref.~~\cite{Becirevic:2023aov} which combines previous lattice QCD results from FNAL/MILC~\cite{Bailey:2015dka} with the recent ones from HPQCD~\cite{Parrott:2022rgu}. Other determinations of the form factors can be found in Ref.~\cite{Gubernari:2023puw}.

For the case of the $B\rightarrow D$ transition we instead use the FLAG results~\cite{FlavourLatticeAveragingGroupFLAG:2021npn} combining FNAL/MILC~\cite{MILC:2015uhg} and HPQCD~\cite{Na:2015kha} results for the vector and scalar form factors. Instead, for the tensor form factor we assume that the ratio $f_T(q^2)/f_+(q^2)$ is constant and take the result from Ref.~\cite{Atoui:2013zza}, where it was found that $f_T(q_0^2)/f_+(q_0^2)=1.08(7)$ at $q_0^2=11.5$~GeV$^2$.

\subsection{$B\rightarrow V$}
Regarding the transition with a vector meson, $V$, in the final state, the matrix elements are given by
\begin{equation}
	\begin{split}
		\big\langle \bar{V}(k)|\bar{q}_i\gamma_{\mu}&(1\mp\gamma_5)b|\bar{B}(p)\big\rangle =  \epsilon_{\mu\nu\rho\sigma}\varepsilon^{*\nu}p^\rho k^\sigma\frac{2V(q^2)}{m_B+m_{V}}\\
		\mp&i\varepsilon^*_{\mu}(m_B+m_{V})A_1(q^2)\\
		\pm&i(p+k)_{\mu}\left(\varepsilon^*\cdot q\right)\frac{A_2(q^2)}{m_B+m_{V}}\\
		\pm&iq_{\mu}\left(\varepsilon^*\cdot q\right)\frac{2m_{V}}{q^2}\left[A_3(q^2)-A_0(q^2)\right]\,,\\
		\big\langle \bar{V}(k)|\bar{q}_i\sigma^{\mu\nu}q_{\nu}&(1\mp\gamma_5)b|\bar{B}(p)\big\rangle=2i\epsilon^{\mu\nu\rho\sigma}\varepsilon^*_{\nu}p_{\rho}k_{\sigma}T_1(q^2)\\
		\pm \big[\varepsilon^{*\mu}&(m_B^2-m_{V}^2)-(p+k)^{\mu}\left(\varepsilon^*\cdot q\right)\big]T_2(q^2)\\
		\pm \big[q^{\mu}-&\frac{q^2}{m_B^2-m_{V}^2}(p+k)^{\mu}\big]\left(\varepsilon^*\cdot q\right)T_3(q^2)\,,
	\end{split}
	\label{eq:Vector_FF_BtoKst}
\end{equation}
where $\varepsilon$ corresponds to the polarization of the vector meson in the final state. At $q^2=0$ one finds the consistency conditions $A_3=A_0$ as well as $T_1=T_2$. Moreover, $A_3$ is related to $A_1$ and $A_2$ as
\begin{equation}
    A_3(q^2)=\frac{m_B+m_{K^*}}{2m_{K^*}}A_1(q^2)-\frac{m_B-m_{K^*}}{2m_{K^*}}A_2(q^2)\,.
\end{equation}

For the $B\rightarrow K^*$ transition we take the form factors from Ref.~\cite{Bharucha:2015bzk}, in which lattice QCD results from Ref.~\cite{Horgan:2013hoa} were combined with light-cone sum rules (LCSR).

Instead, several results from lattice QCD at non-zero recoil exist for $B\rightarrow D^*$~\cite{FermilabLattice:2021cdg,Harrison:2023dzh,Aoki:2023qpa}. We use the FNAL/MILC~\cite{FermilabLattice:2021cdg} ones and note that recent results from HPQCD~\cite{Harrison:2023dzh} find a tension between the BaBar and Belle measurements of the differential decay rate and the lattice results. However, given the negligible impact we find on the $B\rightarrow D^{(*)}\ell_{\alpha}\nu$ process, this issue does not play any role in our analysis.

\subsection{Form factor dependence of $\delta \mathcal{B}_{K^{(*)}}$}
We can parameterise the contributions from NP in terms of appropriate combinations of form factors integrated over the relevant phase space and the WCs. In the particular case in which we can neglect neutrino masses in the final state, we find the following quantities for $B\rightarrow K\nu\nu$:
\begin{equation}
    \begin{split}
       \rho_{K,0}&=\int_{0}^{q^2_{\mathrm{max}}}dq^2\frac{(m_B^2-m_K^2)^2}{(m_b-m_s)^2}q^2\lambda_{BK}^{1/2}f_0^2\,,\\
       \rho_{K,+}&=\int_{0}^{q^2_{\mathrm{max}}}dq^2\lambda_{BK}^{3/2}f_+^2\,,\\
       \rho_{K,T}&=\int_{0}^{q^2_{\mathrm{max}}}dq^2\frac{\lambda_{BK}^{3/2}}{(m_B+m_K)^2}q^2f_T^2\,,
    \end{split}
\end{equation}
where $\lambda_{BK}\equiv\lambda_{BK\sqrt{q^2}}$ and we have $q^2_{\mathrm{max}}=(m_B-m_K)^2$. We then define $\eta_{K}^{S(T)}\equiv \rho_{K,0(T)}/\rho_{K,+}$. 

For the $B\rightarrow K^*\nu\nu$ transition, more form factors are involved, such as we define for the (axial-) vector form factors the following quantities:
\begin{equation}
    \begin{split}
        \rho_{K^*,1}&=\int_{0}^{q^2_{\mathrm{max}}}dq^2 q^2\lambda_{BK^*}^{1/2}A_1^2\,,\\
        \rho_{K^*,V}&=\int_{0}^{q^2_{\mathrm{max}}}dq^2\frac{\lambda_{BK^*}^{3/2}}q^2V^2\,,\\
        \rho_{K^*,12}&=32m_B^2m_{K^*}^2\int_{0}^{q^2_{\mathrm{max}}}dq^2\lambda_{BK^*}^{1/2}A_{12}^2\,,
    \end{split}
\end{equation}
where now $q^2_{\mathrm{max}}=(m_B-m_{K^*})^2$ and we again used the abbreviation $\lambda_{BK^*}\equiv \lambda_{BK^*\sqrt{q^2}}$. $A_{12}$ is related to $A_1$ and $A_2$ as
\begin{equation}
    \begin{split}
        A_{12}(q^2)=&\frac{(m_B+m_{K^*})(m_B^2-m_{K^*}^2-q^2)}{16m_Bm_{K^*}^2}A_1(q^2)\\
        &-\frac{\lambda_{BK^*}}{16m_Bm_{K^*}^2(m_B+m_{K^*})}A_2(q^2)\,.
    \end{split}
\end{equation}
Additionally, for the scalar form factor contribution we have
\begin{equation}
    \rho_{K^*,0}=\int_0^{q^2_{\mathrm{max}}}dq^2\frac{\lambda_{BK^*}^{3/2}}{(m_b+m_s)^2}q^2A_0^2\,,
\end{equation}
and finally for the tensor form factors we can write
\begin{equation}
    \begin{split}
        \rho_{K^*,T}=&\int_0^{q^2_{\mathrm{max}}} dq^2\lambda_{BK^*}^{1/2}\bigg\lbrace\lambda_{BK^*}T_1^2\\
        +&(m_B^2-m_{K^*}^2)^2T_2^2
        +\frac{8m_B^2m_{K^*}^2q^2}{(m_B+m_{K^*})^2} T_{23}^2\bigg\rbrace\,.
    \end{split}
\end{equation}
With these definitions, we finally arrive at
\begin{equation}
    \begin{split}
        \eta_{K^*}^V&\equiv 4\frac{\rho_{K^*,1}+\rho_{K^*,12}}{\rho_{K^*,1}+\rho_{K^*,12}+\rho_{K^*,V}},\\
        \eta_{K^*}^S&\equiv \frac{\rho_{K^*,0}}{\rho_{K^*,1}+\rho_{K^*,12}+\rho_{K^*,V}},\\
        \eta_{K^*}^T&\equiv \frac{\rho_{K^*,T}}{\rho_{K^*,1}+\rho_{K^*,12}+\rho_{K^*,V}}\,,
    \end{split}
\end{equation}
whose numerical values can be found in Eq.~(\ref{eq:values_etas}.

\section{Differential decay rates}
\label{sec:diff_dec_rates}
We devote this appendix to summarizing the results on the differential branching fractions for $B\rightarrow K^{(*)}\nu\nu$ as well as the fraction of longitudinally polarized $K^*$ in the final state, for any neutrino mass. These are based in Ref.~\cite{Gratrex:2015hna} and we find they agree with those from Refs.~\cite{Felkl:2021uxi,Felkl:2023ayn}.

\subsection{$B\rightarrow Kn_in_j$}
We can write the differential branching fraction for the $B\rightarrow Kn_in_j$ decay as
\begin{equation}
    \begin{split}
        &\frac{d^2\mathcal{B}}{dq^2d\cos{\theta_\nu}}=\frac{\tau_{B}}{4}\bigg[G^{(0)}(q^2)\\
        +&G^{(1)}(q^2)\cos{\theta_{\nu}}+\frac{G^{(2)}(q^2)}{2}\left(3\cos^2{\theta_{\nu}}-1\right)\bigg]\,,
    \end{split}
    \label{eq:diff_branching_BtoK}
\end{equation}
where $\tau_{B}$ is the total $B$-meson lifetime. The $q^2$-dependent functions ($G^{(i)}=\mathcal{N}\tilde{G}^{(i)}$) are given in terms of the helicity amplitudes as
\begin{equation}
    \begin{split}
        \tilde{G}^{(0)}=&\bigg(4\omega^{+}_{ij}+\frac{\lambda_{\gamma^*}}{3q^2}\bigg)|h^V|^2
        +\bigg(4\omega^{-}_{ij}+\frac{\lambda_{\gamma^*}}{3q^2}\bigg)|h^A|^2\\
        +&\bigg(4\omega^{-}_{ij}+\frac{\lambda_{\gamma^*}}{q^2}\bigg)|h^S|^2
        +\bigg(4\omega^{+}_{ij}+\frac{\lambda_{\gamma^*}}{q^2}\bigg)|h^P|^2\\
        +&16\bigg(\omega^{+}_{ij}-\frac{\lambda_{\gamma^*}}{12q^2}\bigg)|h^{Tt}|^2
        +8\bigg(\omega^{-}_{ij}-\frac{\lambda_{\gamma^*}}{12q^2}\bigg)|h^{T}|^2\\
        +&16(m_{n_i}E_j+m_{n_j}E_i)\mathrm{Im}[h^Vh^{Tt*}]\\
        +&8\sqrt{2}(m_{n_i}E_j-m_{n_j}E_i)\mathrm{Im}[h^Ah^{T*}]\,,
    \end{split}
\end{equation}
where $\omega^{\pm}_{ij}\equiv E_iE_j\pm m_{n_i}m_{n_j}$. The normalization factor is defined as $\mathcal{N}=G_F^2|\lambda_t|^2\alpha_{\mathrm{em}}^2\lambda_{BK}^{1/2}\lambda_{\gamma^*}^{1/2}/(128\pi^5m_B^3q^2)$. Note also that to ease the notation we have used $\lambda_{\gamma^*}\equiv \lambda_{\sqrt{q^2}n_in_j}$. For the other functions, which do not affect the differential branching fraction once we integrate over the di-neutrino angle, we have
\begin{equation}
    \begin{split}
        \tilde{G}^{(1)}=&-4\lambda_{\gamma^*}^{1/2}\Bigg(\mathrm{Re}\bigg[\frac{m_i+m_j}{\sqrt{q^2}}h^Vh^{S*}\\
        +&\frac{m_{n_i}-m_{n_j}}{\sqrt{q^2}}h^Ah^{P*}\bigg]-\mathrm{Im}\left[2h^{Tt}h^{S*}+\sqrt{2}h^Th^{P*}\right]\Bigg)\,,\\
        \tilde{G}^{(2)}=&-\frac{4\lambda_{\gamma^*}}{3q^2}\left(|h^V|^2+|h^A|^2-2|h^T|^2-4|h^{Tt}|^2\right)\,.
    \end{split}
\end{equation}
The helicity amplitudes, in terms of the form factors from Appendix~\ref{sec:Form_factors} and the relevant WCs from the LEFT lagrangian in Eq.~(\ref{eq:Lag_LEFT}) are given by
\begin{equation}
\begin{split}
    h^{V(A)}=&\frac{\lambda_{BK}^{1/2}}{2\sqrt{q^2}}\left(C_{V(A)}^*+C_{V(A)}^{'*}\right)f_+\,,\\
    h^{S(P)}=&\frac{m_B^2-m_K^2}{2}\bigg(\frac{C^*_{S(P)}+C^{'*}_{S(P)}}{m_b-m_s}\\
    +&\frac{m_{n_i}\mp m_{n_j}}{q^2}\left(C_{V(A)}^*+C_{V(A)}^{'*}\right)\bigg)f_0\,,\\
    h^T=&-i\frac{\lambda_{BK}^{1/2}}{\sqrt{2}(m_B+m_K)}\left(C_T^*-C_T^{'*}\right)f_T\,,\\
    h^{Tt}=&-i\frac{\lambda_{BK}^{1/2}}{2(m_B+m_K)}\left(C_T^*+C_T^{'*}\right)f_T\,.
\end{split}
\end{equation}
Note that in Eq.~(\ref{eq:diff_branching_BtoK}) we need to take into account the possibility to have identical final state particles by dividing the results by a factor of 2.
\subsection{$B\rightarrow K^*n_in_j$}
In the case of the decay into the vector meson, which subsequently decays as $K^*\rightarrow K\pi$, we have instead, after integrating over the solid angle in the di-neutrino system, that the differential branching fraction depends solely on two $q^2$-dependent functions, namely
\begin{equation}
    \begin{split}
    \frac{d^2\mathcal{B}}{dq^2d\cos{\theta_K}}=&\frac{3\tau_{B}}{8}\bigg[G^{0,0}_{0}(q^2)
        +\frac{G^{2,0}_{0}(q^2)}{2}\left(3\cos^2{\theta_{K}}-1\right)\bigg]\,.
    \end{split}
\end{equation}

While the differential branching fraction depends solely on $G^{0,0}_0$, the fraction of longitudinally polarized $K^*$ mesons is given by $F_L=(G^{0,0}_0+G^{2,0}_0)/(3G^{0,0}_0)$. For these functions, factorizing them again into $G^{i,0}_0=\mathcal{N}\tilde{G}^{i,0}_0$, with the overall normalization $\mathcal{N}=G_F^2|\lambda_t|^2\alpha_{\mathrm{em}}^2\lambda^{1/2}_{BK^*}\lambda^{1/2}_{\gamma^*}/(128\pi^5m_{B}^3q^2)$, one finds
\begin{equation}
    \begin{split}
        \tilde{G}^{0,0}_0&=\frac{4}{9}\left(3E_iE_j+\frac{\lambda_{\gamma^*}}{4q^2}\right)\sum_{\eta={\pm,0}}\left(|H^V_{\eta}|^2+|H^A_{\eta}|^2\right)\\
        +&\frac{4m_{n_i}m_{n_j}}{3}\sum_{\eta=\pm,0}\left(|H^V_{\eta}|^2-|H^A_{\eta}|^2\right)\\
        +&\frac{4}{3}\bigg( \omega^{-}_{ij}+\frac{\lambda_{\gamma^*}}{4q^2}\bigg)|H^S|^2
        +\frac{4}{3}\bigg(\omega^{+}_{ij}
        +\frac{\lambda_{\gamma^*}}{4q^2}\bigg)|H^P|^2\\
        +&\frac{16}{9}\bigg(3\omega^{+}_{ij}-\frac{\lambda_{\gamma^*}}{4q^2}\bigg)
        \sum_{\eta=\pm,0}|H^{Tt}_{\eta}|^2\\
        +&\frac{8}{9}\bigg(3\omega^{-}_{ij}-\frac{\lambda_{\gamma^*}}{4q^2}\bigg)
        \sum_{\eta=\pm,0}|H^T_{\eta}|^2\\+&\frac{16}{3}(m_{n_i}E_j+m_{n_j}E_i)\mathrm{Im}(\sum_{\eta=\pm,0}H^V_{\eta}H^{Tt*}_{\eta})\\
        +&\frac{8\sqrt{2}}{3}(m_{n_i}E_j-m_{n_j}E_i)\mathrm{Im}(\sum_{\eta=\pm,0}H^A_{\eta}H^{T*}_{\eta})\,,
    \end{split}
\end{equation}
and
\begin{equation}
    \begin{split}
        \tilde{G}^{2,0}_0&=-\frac{4}{9}\left(3E_iE_j+\frac{\lambda_{\gamma^*}}{4q^2}\right)\bigg[\sum_{\eta={\pm}}\left(|H^V_{\eta}|^2+|H^A_{\eta}|^2\right)\\
        -&2|H_0^V|^2-2|H_0^A|^2\bigg]-\frac{4m_{n_i}m_{n_j}}{3}\bigg[\sum_{\eta=\pm}\big(|H^V_{\eta}|^2\\
        -&|H^A_{\eta}|^2\big)-2|H_0^V|^2+2|H_0^A|^2\bigg]\\
        +&\frac{8}{3}\bigg( \omega^{-}_{ij}+\frac{\lambda_{\gamma^*}}{4q^2}\bigg)|H^S|^2
        +\frac{8}{3}\bigg(\omega^{+}_{ij}
        +\frac{\lambda_{\gamma^*}}{4q^2}\bigg)|H^P|^2\\
        -&\frac{16}{9}\bigg(3\omega^{+}_{ij}-\frac{\lambda_{\gamma^*}}{4q^2}\bigg)
        \bigg[\sum_{\eta=\pm}|H^{Tt}_{\eta}|^2-2|H^{Tt}_0|^2\bigg]\\
        -&\frac{8}{9}\bigg(3\omega^{-}_{ij}
        -\frac{\lambda_{\gamma^*}}{4q^2}\bigg)\bigg[\sum_{\eta=\pm}|H^T_{\eta}|^2-2|H^T_0|^2\bigg]\\
        -&\frac{16}{3}(m_{n_i}E_j+m_{n_j}E_i)
        \mathrm{Im}(\sum_{\eta=\pm}H^V_{\eta}H^{Tt*}_{\eta}-2H^V_0H^{Tt*}_0)\\
        -&\frac{8\sqrt{2}}{3}(m_{n_i}E_j
        -m_{n_j}E_i)\mathrm{Im}(\sum_{\eta=\pm}H^A_{\eta}H^{T*}_{\eta}-2H^A_0H^{T*}_0)\,.
    \end{split}
\end{equation}
The necessary helicity amplitudes are given now by
\begin{equation}
    \begin{split}
        H^{V(A)}_0=&\frac{4im_Bm_{K^*}}{\sqrt{q^2}}(C_{V(A)}^*-C_{V(A)}^{'*})A_{12}\,,\\
        H^{V(A)}_{\pm}=&\frac{i}{2(m_B+m_{K^*})}\bigg[\pm (C_{V(A)}^*+C_{V(A)}^{'*})\lambda_{BK^*}^{1/2}V\\
        -&(m_B+m_{K^*})^2(C_{V(A)}^*-C_{V(A)}^{'*})A_{1}\bigg]\,,\\
        H^{S(P)}=&\frac{i\lambda_{BK^*}^{1/2}}{2}\bigg(\frac{C_{S(P)}^*-C_{S(P)}^{'*}}{(m_b+m_s)}\\
        +&\frac{m_{n_i}\mp m_{n_j}}{q^2}(C_{V(A)}^*-C_{V(A)}^{'*})\bigg)A_0\,,\\
        H^T_0=&\frac{2\sqrt{2}m_Bm_{K^*}}{m_B+m_{K^*}}(C_T^*+C_T^{'*})T_{23}\,,\\
        H^{Tt}_0=&\frac{2m_Bm_{K^*}}{m_B+m_{K^*}}(C_T^*-C_T^{'*})T_{23}\,,\\
        H^T_{\pm}=&\frac{1}{\sqrt{2q^2}}\bigg(\pm(C_T^*-C_T^{'*})\lambda_{BK^*}^{1/2}T_1\\
        -&(C^*_T+C_T^{'*})(m_B^2-m_{K^*}^2)T_2\bigg)\,,\\
        H^{Tt}_{\pm}=&\frac{1}{2\sqrt{q^2}}\bigg(\pm(C_T^*+C_T^{'*})\lambda_{BK^*}^{1/2}T_1\\
        -&(C^*_T-C_T^{'*})(m_B^2-m_{K^*}^2)T_2\bigg)\,.
    \end{split}
\end{equation}

\bibliographystyle{JHEP}
\bibliography{Biblio}
\end{document}